\documentclass[a4paper]{aa} %
\usepackage{graphicx}         
\usepackage{txfonts}
\usepackage{natbib}
\usepackage{amsfonts}
\usepackage{threeparttable}  
\usepackage{booktabs}
\usepackage{wasysym}
\newcommand{\degree}{\ensuremath{^\circ}}  
\newcommand{\source}{\object{HD141569A}}
\newcommand{\CII}{\ion{C}{II}}
\newcommand{\OI}{\ion{O}{I}}
\newcommand{\CIIfs}{[\ion{C}{II}]}
\newcommand{\OIfs}{[\ion{O}{I}]}

\begin{document}
\title{Gas lines from the 5-Myr old optically thin disk around
  \source\thanks{Based on observations made with ESO Telescopes at the La Silla Paranal Observatory under programme ID 079.C-0602(A)}}

\subtitle{Herschel observations and modeling \thanks{Herschel is an
    ESA space observatory with science instruments provided by
    Principal Investigator consortia. It is open for proposals for
    observing time from the worldwide astronomical community.}}

   \author{
     W.-F. Thi\inst{1}, 
     C. Pinte\inst{1},
     E. Pantin\inst{2},
     J.C. Augereau\inst{1},
     G. Meeus\inst{3},
     F. M\'{e}nard\inst{1,4},
     C. Martin-Za\"{i}di\inst{1},
     P. Woitke\inst{5},
     P. Riviere-Marichalar\inst{6}
     I. Kamp\inst{6},
     A. Carmona\inst{1},
     G. Sandell\inst{7},
     C. Eiroa\inst{3},
     W. Dent\inst{8},
     B. Montesinos\inst{3},
     G. Aresu\inst{6},
     R. Meijerink\inst{6},
     M. Spaans\inst{6},
     G. White\inst{9,10},
     D. Ardila\inst{11},
     J. Lebreton\inst{1},
     I. Mendigut\'{i}a\inst{12},
     S. Brittain\inst{12}
}

\institute{
  UJF-Grenoble 1 / CNRS-INSU, Institut de Plan\'{e}tologie et d'Astrophysique (IPAG) UMR 5274, Grenoble, F-38041, France\\
  \email{Wing-Fai.Thi@obs.ujf-grenoble.fr} \and Laboratoire AIM,
  CEA/DSM - CNRS - Universit\'{e} Paris Diderot, IRFU/SAP, F-91191 sur
  Yvette, France \and 
  Dep. de F\'{i}sica Te\'{o}rica, Fac. de Ciencias, UAM Campus
  Cantoblanco, 28049 Madrid, Spain \and UMI -- LFCA, CNRS / INSU
  France, and Dept. de Astronomia y Obs. Astronomico Nacional,
  Universidad de Chile, Casilla 36-D, Correo Central, Santiago, Chile
  (UMI 3386) \and 
  SUPA, School of Physics \& Astronomy, University of St.~Andrews,
  North Haugh, St.~Andrews KY16 9SS, UK \and 
  Kapteyn Astronomical Institute, P.O. Box 800, 9700 AV Groningen, The
  Netherlands \and SOFIA-USRA, NASA Ames Research Center, Mail Stop
  N211-3, Building N211/Rm. 249, Moffett Field, CA 94035, USA \and
  ALMA, Avda Apoquindo 3846, Piso 19, Edificio Alsacia, Las Condes,
  Santiago, Chile \and Astrophysics Group, Department of Physics \&
  Astronomy, The Open University, UK \and RAL Space, The Rutherford
  Appleton Laboratory, Didcot, Oxfordshire, UK \and NASA Herschel
  Science Center, California Institute of Technology, Mail Code
  100-22, Pasadena, CA 91125, USA \and Department of Physics and
  Astronomy, 118 Kinard Laboratory, Clemson University, Clemson, SC
  29634, USA}
\authorrunning{Thi et al.}  
\titlerunning{Herschel observations and
  modeling of \object{HD~141569A}}

   \date{Received 2013; accepted 10 september 2013}

 
  \abstract
  {The gas- and dust dissipation processes in disks around young stars
    remain uncertain despite numerous studies. At the distance of
    $\sim$~99--116\,pc, \source\ is one of the nearest HerbigAe stars
    that is surrounded by a tenuous disk, probably in transition
    between a massive primordial disk and a debris disk. Atomic and
    molecular gases haves been found in the structured 5-Myr old
    \source\ disk, making \source\ the perfect object within which to
    directly study the gaseous atomic and molecular component.}
  {We wish to constrain the gas and dust mass in the disk around
    \source.}
  {We observed the fine-structure lines of \OI\ at 63 and 145 $\mu$m
    and the \CII\ line at 157~$\mu$m with the {{\it PACS}} instrument
    onboard the {{\it Herschel Space Telescope}} as part of the
    open-time large programme {{\sc GASPS}}. We complemented the
    atomic line observations with archival {{\it Spitzer}}
    spectroscopic and photometric continuum data, a ground-based {{\it
        VLT-VISIR}} image at 8.6 $\mu$m, and $^{12}$CO fundamental
    ro-vibrational and pure rotational $J$=3--2 observations.  We
    simultaneously modeled the continuum emission and the line fluxes
    with the Monte Carlo radiative transfer code {{\sc MCFOST}} and
    the thermo-chemical code {{\sc ProDiMo}} to derive the disk gas-
    and dust properties assuming no dust settling.}
  { The models suggest that the oxygen lines are emitted from the
    inner disk around \source, whereas the \CIIfs\ line emission is
    more extended. The CO submillimeter flux is emitted mostly by the
    outer disk.  Simultaneous modeling of the photometric and line
    data using a realistic disk structure suggests a dust mass derived
    from grains with a radius smaller than 1~mm of $\sim$ 2.1 $\times$
    10$^{-7}$ M$_{\odot}$ and from grains with a radius of up to 1 cm
    of 4.9 $\times$ 10$^{-6}$ M$_{\odot}$. We constrained the
    polycyclic aromatic hydrocarbons (PAH) mass to be between
    2$\times$10$^{-11}$ and 1.4$\times$10$^{-10}$ M$_{\odot}$ assuming
    circumcircumcoronene (C$_{150}$H$_{30}$) as the representative
    PAH.  The associated PAH abundance relative to hydrogen is lower
    than those found in the interstellar medium (3$\times$10$^{-7}$)
    by two to three orders of magnitude. The disk around \source\ is
    less massive in gas (2.5 to 4.9 $\times$ 10$^{-4}$ M$_{\odot}$ or
    67 to 164 M$_{\oplus}$) and has a flat opening angle ($<$ 10\%).}
  {We constrained simultaneously the silicate dust grain, PAH, and gas
    mass in a $\sim$5-Myr old Herbig~Ae disk. The disk-averaged
    gas-to-dust-mass is most likely around 100, which is the assumed
    value at the disk formation despite the uncertainties due to
    disagreements between the different gas tracers. If the disk was
    originally massive, the gas and the dust would have
    dissipated at the same rate.}

   
   \maketitle
%

\section{Introduction}
 
Constraining the gas- and dust mass in the disks around young stars is
paramount for our understanding of the planet formation process (e.g.,
\citealt[][]{Armitage2010apf}). The determination of the dust mass can
be routinely made by fitting the spectral energy distribution (SED)
and images of disks. Sophisticated and reliable radiative-transfer
codes are available (e.g.,
\citealt[][]{Pinte2006A&A...459..797P,Pinte2009A&A...498..967P,Min2009A&A...497..155M,Dullemond2004A&A...417..159D}). The
accuracy of the dust mass estimates is limited by the uncertainties in
the opacities due to our limited knowledge of the grain composition,
the shape, and the size distribution.  However, up to now, the
estimates of the solid mass in disks are much more accurate than
gas-mass estimates, so that most studies assume the interstellar
gas-to-dust-mass ratio of 100 remains constant throughout the disk
lifetime. The most common method involves the observation of
rotational emission lines from CO and its isotopologues $^{13}$CO and
C$^{18}$O. The premise of this method is that the typical H$_2$/CO
conversion factor for dense clouds, $\sim$~10$^{-4}$ for $^{12}$CO,
remains valid for protoplanetary disks.  Most disk gas-mass estimates
using this method show that the gas-to-dust ratios in disks may be
lower than 100.  The reasons for the small disk-gas masses are thought
to be CO freeze-out on grain surfaces and/or CO photodissociation,
especially for disks with masses lower than 10$^{-3}$ M$_\odot$.

The {\sc PACS} instrument \citep{Poglitsch2010} onboard the Herschel
Space Telescope \citep{Pilbratt2010} is sensitive enough to observe
lines from species that result from the photodissociation of CO
(atomic oxygen and singly ionized carbon) in disks. We used
observations of the O and C$^+$ fine-structure emissions in addition
to CO and, when available, $^{13}$CO data to constrain accurately the
gas mass around the classical T~Tauri stars \object{TW~Hya}
\citep{thi2010A&A...518L.125T} and \object{ET~Cha}
\citep{Woitke2011A&A...534A..44W}, as well as around the HerbigAe star
\object{HD~169142} \citep{Meeus2010A&A...518L.124M} and
\object{HD~163296} \citep{Tilling2012A&A...538A..20T}. These studies
are part of the large open-time program {\em GASPS}
\citep{Dent2013PASP..125..477D,Mathews2010A&A...518L.127M,Pinte2010A&A...518L.126P}. The
disk gas-to-dust-mass ratio may vary by orders of magnitudes between
objects. The gas disk around \object{TW~Hya} is ten times less massive
than the total amount of solids up to 10 cm in radius than in the
interstellar medium where the maximum radius is below one micron while
the gas remains 100 times more massive in the disk around
\object{HD~169142}, and over 1000 times more massive in the disk
around \object{ET~Cha}.

In this paper, we extend our study of disk dust and gas masses to the
low-mass optically thin disk around \source\ as a part of the
Herschel-GASPS programme.  The use of atomic and molecular lines as
gas mass tracers has been explored by \citet{Kamp2010A&A...510A..18K}.

At a distance of $\sim$~99~pc \citep{vandenancker1997A&A...324L..33V}
to 116.1$\pm$8.1 \citep{Merin2004A&A...419..301M}, \source\ is one the
nearest HerbigAe stars and has an estimated age of 4.7 $\pm$ 0.3~Myr
\citep{Merin2004A&A...419..301M}. \source\ lies at the edge of the
high Galactic latitude core \object{MBM37}
\citep{Caillault1995ApJ...441..261C}. The disk around \source\ was
first detected from its excess emission over the photospheric flux in
the mid-infrared data taken by the {\it Infrared Astronomical
  Satellite (IRAS)}
\citep{Andrillat1990A&A...233..474A,Walker1988PASP..100.1509W,Sylvester1996MNRAS.279..915S}. Subsequent
high-resolution scattered-light imagings with the {\it Hubble Space
  Telescope} showed that \source\ is surrounded by a complex ring
structure with a non-monotonic decreasing surface brightness with
radius, instead of a uniformly disk
\citep{Augereau1999A&A...350L..51A,Weinberger1999ApJ...525L..53W}.
Further coronographic high-quality observations revealed a strong
brightness asymmetry interpreted as the signature of a massive planet
embedded within the disk
\citep{Mouillet2001A&A...372L..61M,Boccaletti2003ApJ...585..494B,Wyatt2005A&A...440..937W}
and two spiral arms that may be due to the companion star
\citep{Clampin2003AJ....126..385C}.  The observations constrain the
inner ring peak emission at $\sim$200~$\pm$~5~AU and a width of
$\sim$50~AU. Direct and coronographic imaging studies at optical and
near-infrared wavelengths are not sensitive to material within
100~AU. Ground-based mid-infrared imagings at 10.8 and 18.2 $\mu$m
\citep{Fisher2000ApJ...532L.141F} and at 17.9 and 20.8~$\mu$m
\citep{Marsh2002ApJ...573..425M} support the presence of warm dust
grains within 100~AU to the star, as predicted by
\citet{Augereau1999A&A...350L..51A} to explain the mid-infrared
spectral energy distribution.  Fits to the SED enable the solid disk
mass to be estimated as $6.7\times 10^{-6}$ M$_\odot$ or 2.2
M$_\oplus$ \citep{Li2003ApJ...594..987L}.  The fit to the SED also
suggests that grains have grown to $\sim$~10~cm. PAH emissions have
been detected toward \source\ by the {\sc Spitzer} Space Telescope
\citep{Keller2008ApJ...684..411K}.

\source\ shows a rich CO ro-vibrational ($v \ge$1, $\Delta v$=1)
emission spectrum around 4.65 $\mu$m
\citep{Brittain2003ApJ...588..535B,Goto2006ApJ...652..758G,Brittain2007ApJ...659..685B,Salyk2011ApJ...743..112S},
testifying of the presence of warm molecular gas. The spatially
resolved data showed that CO emission arises from beyond 11 $\pm$ 2~AU
\citep{Goto2006ApJ...652..758G}. No companion more massive than 22
Jupiter mass at 7~AU has been found by the sparse aperture-masking
technique \citep{Lacour2011A&A...532A..72L}.

The fine-structure line emission has been studied by
\citet{Jonkheid2006A&A...453..163J} using a 1+1D chemo-physical
code. They made predictions for the \OIfs\ 63 and 145 $\mu$m and
\CIIfs\ 157 $\mu$m lines based on a model that matched the $^{12}$CO
$J$=3-2 observations by \citet{Dent2005MNRAS.359..663D} using the {\it
  James Clerk Maxwell Telescope
  (JCMT)}. \citet{Sandell2011ApJ...727...26S} observed the continuum
emission at 450~$\mu$m and 850~$\mu$m with {\it SCUBA} at the {\it
  JCMT}. In this paper, we use the continuum Monte Carlo code {\sc
  MCFOST} to model the continuum photometry and images
\citep{Pinte2006A&A...459..797P}, and the chemo-physical code {\sc
  ProDiMo}, which self-consistently computes the chemistry and gas
cooling of disks
\citep{Woitke2009A&A...501..383W,thi2011MNRAS.412..711T,thi2013A&A...551A..49T}
to interpret the {\it Herschel} and {\it JCMT} line data.

The paper is organized as follows: the {\it Herschel} data are shown
in Sect.~\ref{observations_first_analysis} where a simple analysis is
also provided. In Sect.~\ref{ISM_emission}, we discuss the possible
contribution of interstellar emission to the observed fine-structure
fluxes. A modeling of the continuum with {\sc MCFOST} and the gas
lines with {\sc ProDiMo} is presented in
Sect.~\ref{continuum_modeling} and ~\ref{gas_modeling}. The derived
solid and gas masses in \source\ are discussed in
Sect.~\ref{diskussion}, followed by the conclusions in
Sect.~\ref{conclusion}.


\begin{table}
\begin{center}
  \caption{PACS photometry with 1~$\sigma$ shot-noise
    error. Calibration uncertainties are 5\% and 10\% in the blue, and
    red filter respectively.}\label{PAC_photmetry}
		\begin{tabular}{lll}
                  \toprule
\noalign{\smallskip}   
                  Band & $\lambda$ ($\mu$m) & Flux (Jy)\\
\noalign{\smallskip}   
\hline
\noalign{\smallskip}   
                   blue    & 70 & 3.09 $\pm$ 0.11\\
                   red      & 160 & 1.33 $\pm$ 0.05\\

\noalign{\smallskip}     
\bottomrule
\end{tabular} 
\end{center}
\end{table}

\begin{table*}
\begin{center}
\caption{Lines observed by {\em Herschel-PACS} \citep{Meeus2012A&A...544A..78M}. The errors are 1$\sigma$ and the upper limits are 3$\sigma$. The calibration error adds an extra $\sim$~30\% uncertainty. The CO data are taken from \citet{Dent2005MNRAS.359..663D} and have uncertainties of $\sim$~30\%. n. a. means not available.}
\label{table_results} 
\begin{tabular}{lr@{.}lr@{~$\pm$~}lr@{}r@{.}l}   
  \hline 
  \noalign{\smallskip}   
  \multicolumn{1}{c}{Line} & \multicolumn{2}{c}{$\lambda$} & \multicolumn{2}{c}{Cont. flux} & \multicolumn{3}{c}{Line flux}\\
  & \multicolumn{2}{c}{($\mu$m)}  &  \multicolumn{2}{c}{(Jy)}& \multicolumn{3}{c}{(10$^{-18}$ W m$^{-2}$)}\\  
  \noalign{\smallskip} 
  \hline
  \noalign{\smallskip} 
  \OIfs\  $^3$P$_1 \rightarrow ^3$P$_2$ & 63&183 & 2.98&0.01 & & 245&3~$\pm$~4.8\\
  o-H$_2$O $8_{18}-7_{07}$      &63&324 & 2.98&0.01 & $<$ & 8&6\\
  CO $J$=36 $\rightarrow$ 35  & 72&85 & 3.91&0.03 & $<$ & 10&6\\
  CO $J$=33 $\rightarrow$ 32  & 76&36 & 3.30&0.03 & $<$ &10&4\\
  p-H$_2$O $3_{22}-2_{11}$ & 89&99 & 2.89&0.03      & $<$ &  7&6\\
  CO $J$=29 $\rightarrow$ 28  & 90&16 & 2.77&0.02 & $<$ &5&9\\
  CO $J$=18 $\rightarrow$ 17  & 144&78 & 1.09&0.07 & $<$ & 21&5\\
  \OIfs $^3$P$_0 \rightarrow ^3$P$_1$ & 145&525 &  1.29&0.01 & & 24&9~$\pm$~1.4   \\
  \CIIfs\  $^2$P$_{3/2} \rightarrow ^2$P$_{1/2}$ &  157&74 & 1.18&0.02  &  & 11&4~$\pm$~1.8\\
  o-H$_2$O $2_{12}-1_{01}$ & 179&52 & 0.85&0.03 & $<$ & 6&0\\
  o-H$_2$O $2_{21}-2_{12}$ & 180&42 & 0.82&0.04 & $<$ & 6&9\\
  SO$_2$ $27_{11,17}-26_{10,16}$$^a$ & 189&67 &  0.50&0.10 & &  5&0~$\pm$~0.7\\ 
  \noalign{\smallskip}
  \hline
  \noalign{\smallskip} 
$^{12}$CO $J$=3--2 & 866&96 & \multicolumn{1}{c}{n.\ a.} & & & 0&1~$\pm$~0.008 \\
\noalign{\smallskip} 
\hline
\end{tabular}
\end{center}  
$^a$ Tentative detection and assignment. The line is probably an artifact.\\
\end{table*}
\normalsize 

\section{Herschel and VISIR observations and first analysis}\label{observations_first_analysis}

\begin{figure*}[!ht]
\centering
\resizebox{\hsize}{!}{\includegraphics[angle=90]{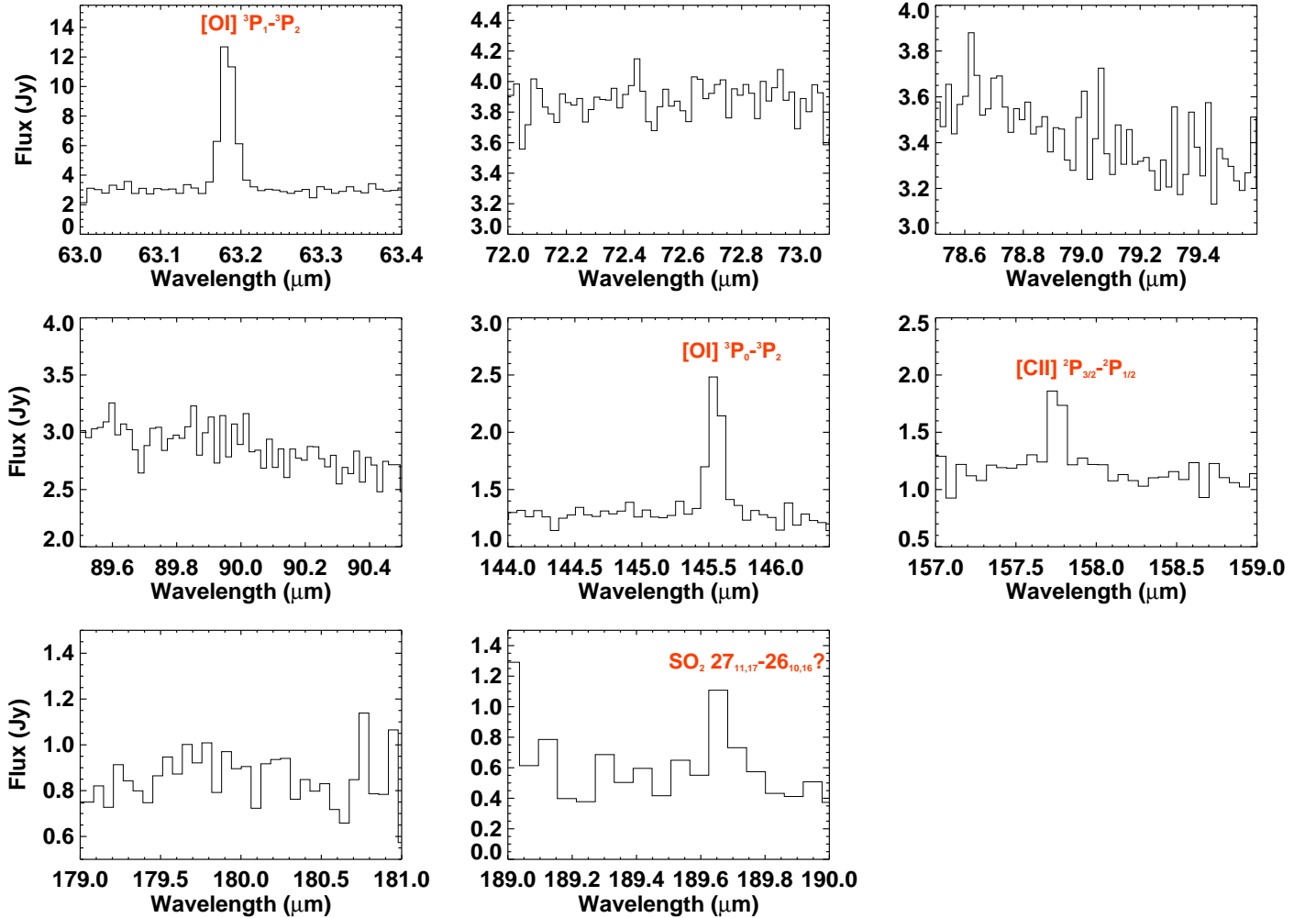}}
\caption{Herschel spectra toward \source.The peak at 189.7,
    which may correspond to a SO$_2$ line is most likely an
    artifact.}
  \label{fig_herschel_spectra}          
\end{figure*}  

\subsection{Observations}\label{observations}

For \source\ we obtained photometry in the blue (70 $\mu$m) and red
(160 $\mu$m) band (obsid 2342215382 and 1342215383) of the PACS camera
by assembling mini-scan scan maps with a scan speed of $20'' s^{-1}$
and a scan length of $2'$. The total duration of the scan for this map
was 731\,sec, with an on-source time of 146\,seconds. The results of
the photometry are given in Table ~\ref{PAC_photmetry} and have an
absolute accuracy estimated to be 5\% for the blue channel and 10\%
for the the red channel.  These values agree very well with the
observed IRAS flux densities and also with the continuum flux
densities measured with the PACS spectrometer
(Table~\ref{table_results}). We also observed with the PACS
spectrometer to target the \OIfs\ line at 63 $\mu$m in line scan mode,
the \OIfs and the \CIIfs\ lines at 145 and 158 $\mu$m, respectively in
range scan mode (obsid 1342190376 PacsLineSpec and obsid 1342190375
and 1342204340 (D3) PacsRangeSpec). We reduced the spectra with the
standard packages in HIPE 7.0, which
include flat-field correction.\\
\indent We detected the two \OIfs\ and the \CIIfs\ lines; see
Table~\ref{table_results}.  The absolute accuracy of PACS spectroscopy
is currently estimated to be about 30\%, but is expected to improve in
the future as more calibration data are acquired and our understanding
of the instrument behavior improves (see Herschel datareduction HIPE
manual). Figure~\ref{fig_herschel_spectra} shows the {\it Herschel}
spectra. We may have detected an emission line at 189.67 microns,
which we assign tentatively to the SO$_2$ $27_{11,17}-26_{10,16}$
transition. Upper limits for transitions of CO and water are given in
Table~\ref{table_results} and have previously been discussed in
\citet{Meeus2012A&A...544A..78M}. The \CIIfs\ at 147 $\mu$m toward
\source\ was detected by \citet{Lorenzetti2002A&A...395..637L} using
{\it ISO-LWS} with a flux of 3.8$\pm$0.5 $\times$10$^{-16}$ W m$^{-2}$
with a beam size of $\sim$~80\arcsec, much higher than the value found
by {\it Herschel-PACS} (11.4~$\times$~10$^{-18}$ W m$^{-2}$).

\begin{figure*}[!ht]
\centering
\includegraphics[scale=0.5,angle=0]{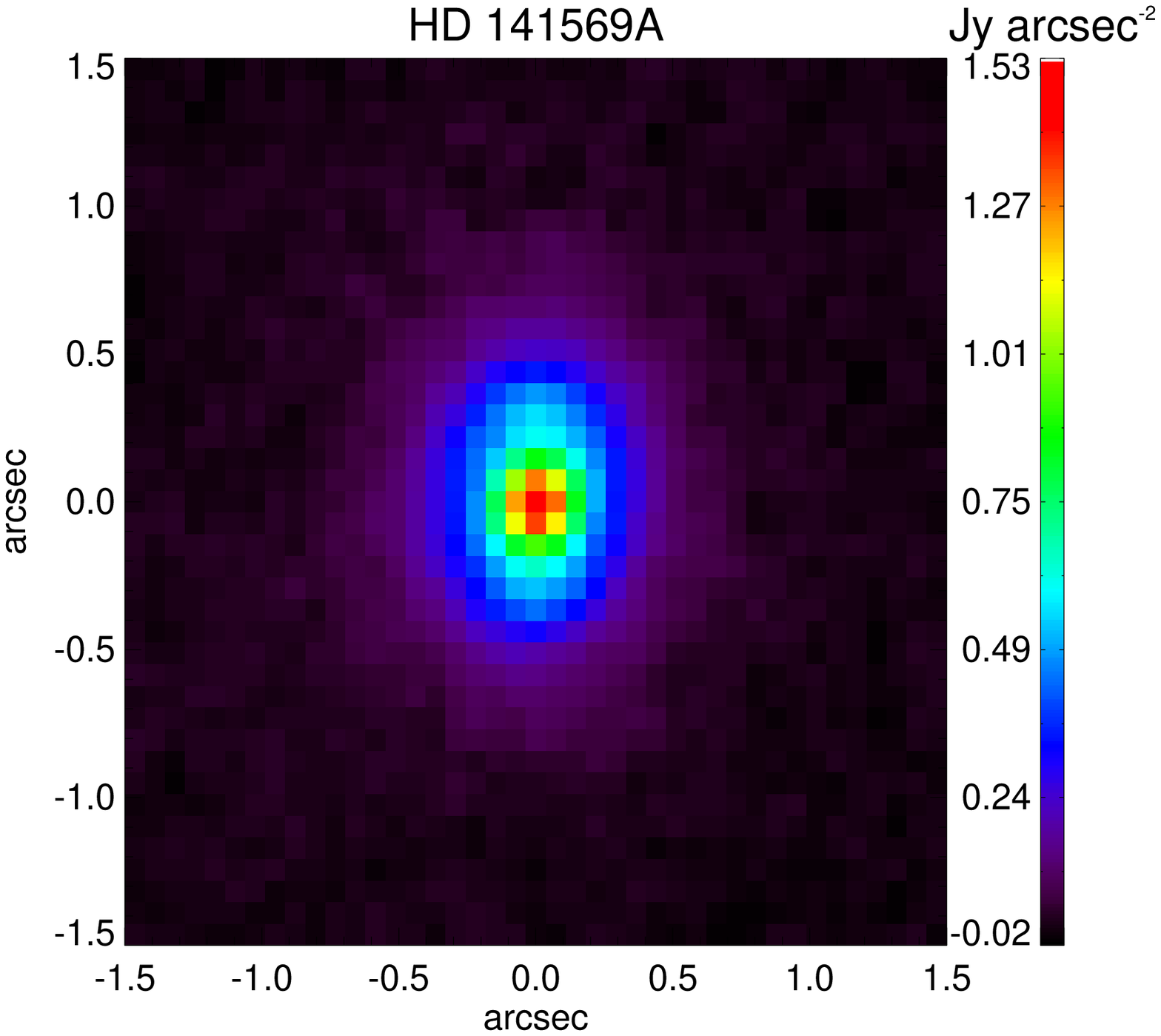}
\includegraphics[scale=0.5,angle=0]{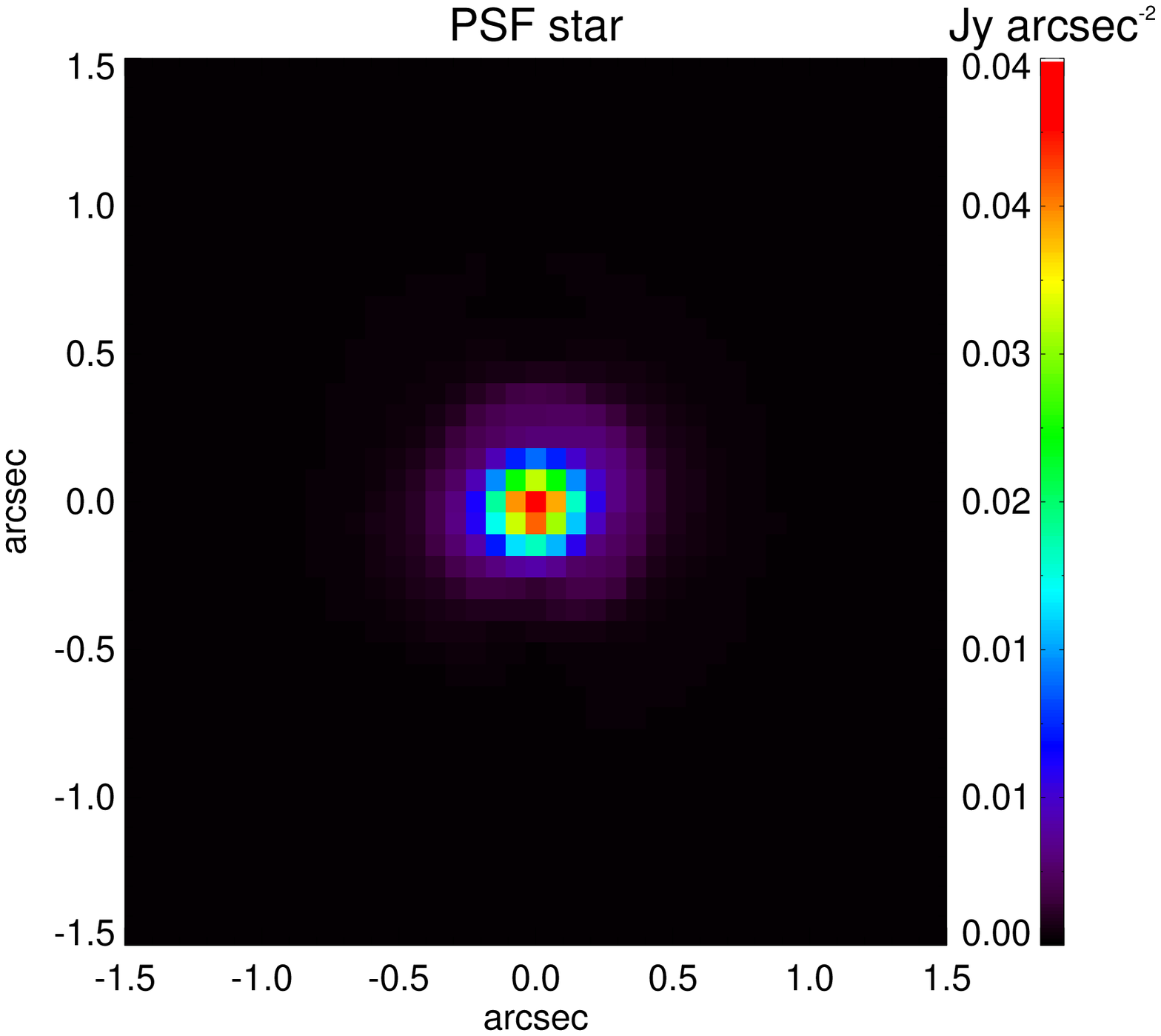}
\caption{Observations at the VLT with the VISIR instrument and the 8.6
  microns filter (up is North, right is West). The left panel
  shows the image of \object{HD~141569A} and the right panel the
  point-spread-function star \object{HD~146791}. The disk around
  \object{HD~141569A} is well resolved along its major axis in the
  North-South direction.}
\label{fig_VISIR_images}          
\end{figure*}

\source\ was observed on 26 March 2005 with the {\it VLT/VISIR}
instrument, the VLT Imager and Spectrometer for mid-IR
\citep{Lagage2004Msngr.117...12L}.  The sensitivity monitored using
standard stars was good (4 mJy/10$\sigma$/1h). The seeing was fair
(0.7\arcsec in the optical range).  The quadrangular chop-nod scheme
was used to preserve the best angular resolution.  The four beams were
registered offline using a dedicated tool.  The total on-source
observing time was 533 s.  We used the filter PAH1 ($\lambda_c=8.6\
\mu m$, $\Delta\lambda=0.42\ \mu m$), centered on the PAH emission
feature.  The integrated photometry is 0.532 $\pm$ 0.05 Jy.  The
standard star \source\ was observed shortly before \source. It was
used as an estimate of the point spread function (PSF).  Surface
brightness profiles with and without subtracting the 

star were extracted along the major axis of the disk at a position
angle of 0 degrees and were compared to simulated ones obtained by
convolving images models with the PSF (followed by PSF
subtraction). The PAH emission from \object{HD~141569A} is extended in
the north-south direction (Fig.~\ref{fig_VISIR_images}). Finally, we
complemented the photometric points with archival data (see
Table~\ref{table_photometry}) and CO ro-vibrational observations
\citep{Brittain2007ApJ...659..685B}, and we list in
Table~\ref{table_results} the CO $J$=3--2 line flux from
\citet{Dent2005MNRAS.359..663D}.

\subsection{Interstellar material toward
  \source}\label{ISM_emission}

Two high Galactic latitude clouds (\object{MBM37} and the Lynds dark
cloud \object{L134N}) with a hydrogen molecular fraction of 0.4 (a
value of 0.5 indicates a fully molecular gas) seem to be located along
the line-of-sight of \source\ \citep{Sahu1998ApJ...504..522S}. The
emissions from these clouds may contaminate the disk line
fluxes. \citet{Sahu1998ApJ...504..522S} found a total HI column
density of $N$(\ion{H}{I})=8.06~$\pm$~0.08~$\times$~10$^{20}$
cm$^{-2}$ (including the two velocity components) and a H$_2$ column
density of $N$(H$_2$)=1.51 $\times$ 10$^{20}$~$\pm$~9.0 $\times$
10$^{19}$ cm$^{-2}$ from their CH observations, whose velocity at 0 km
s$^{-1}$ is close to the CO
velocities. \citet{Penprase1992ApJS...83..273P} derived a reddening
$E(B-V)=0.10\pm 0.03$ for \source.

\source\ lies just outside the CO contour of \object{MBM37}, the
molecular core associated with \object{L134N}
\citep{Caillault1995ApJ...441..261C}. Diffuse and molecular clouds can
emit the oxygen- and carbon fine-structure emissions detected in the
large pixel of PACS (9 \arcsec~$\times$~9\arcsec), especially the
\CIIfs\ transition, which has a low critical density of
2.8~$\times$~10$^{3}$ cm$^{-3}$, whereas the critical density for the
\OIfs\ 63 and 145 $\mu$m lines are 4.7~$\times$~10$^5$ and
1.0~$\times$10$^5$~cm$^{-3}$, respectively (see
Table~\ref{line_parameters}). The critical density is the minimum gas
density for the upper level of a transition to be populated in
LTE. Lines are predominantly emitted in regions with a similar density
or one that is higher than the critical densities.

\citet{Martin2008A&A...484..225M} observed
interstellar molecular hydrogen absorption lines in the UV using the
{\it FUSE} satellite and found a column of 2.1$^{+1.2}_{-0.8}$
$\times$ 10$^{20}$~cm$^{-2}$.

The total hydrogen column density of \object{MBM37} suggests the
possible contamination of the \source\ disk emission by diffuse
photodissociation region emission that fills the beam. We
  estimated the emission from a 1D photodissociation region
  (e.g., \citealt[][]{Hollenbach1999RvMP...71..173H}). According to the
    models of \citet{Kaufman99}, a cloud irradiated by the standard
    interstellar UV field (Draine field $G_o$=1) and with density of
  10$^2$-10$^3$ cm$^{-3}$ has a surface brightness of
  10$^{-8}$-10$^{-9}$ W m$^{-2}$ sr$^{-1}$ in \CIIfs. The solid angle
  for a pixel size of 9 \arcsec $\times$ 9\arcsec\ is 1.90 $\times$
  10$^{-9}$ sr. A cloud that fills the pixel will emit (2--20)
  $\times$ 10$^{-18}$ W m$^{-2}$ in \CIIfs, $\sim$ 2 $\times$
  10$^{-19}$ W m$^{-2}$ in \OIfs\ 63~$\mu$m, and $\sim$ 2 $\times$
  10$^{-20}$ W m$^{-2}$ in \OIfs\ 145~$\mu$m.

  Alternatively, we can use the flux in the large {\it ISO-LWS} beam
  of $\sim$~80\arcsec and estimate the flux in the {\it Herschel}
  beam. We found a value of 4.8$\times$ 10$^{-18}$ W m$^{-2}$ in the
  {\it Herschel} beam assuming a uniform emitting area, a factor two
  lower than the observed flux. Therefore part of detected \CIIfs\
  flux in the {\it Herschel} beam may be emitted from a
  foreground/background cloud.

Theoretically, cloud emission cannot account for the \OIfs\ fluxes, but
it is possible that part of the \CIIfs\ emission is emitted from
the clouds associated with \source\ or from diffuse clouds in the
line-of-sight. However, if the cloud is more extended than the chop
positions, any large-scale emission will be cancelled out. We also
checked for extended emission from cloud or outflow emission in the
25 PACS pixels (see Fig.~\ref{OI63_spaxels}, \ref{OI145_spaxels}, and
\ref{CII_spaxels}). Because there is no evidence for extended emission
along the line-of-sight of \source, we assumed that the entire \OIfs\
fluxes arise from the circumstellar disk, while the \CIIfs\ flux may
have contributions from interstellar and circumstellar material.

\subsection{Simple disk analysis of Herschel line
  fluxes}\label{simple_analysis}
\begin{table}
\centering
\caption{Molecular data of the lines.}
\label{line_parameters}
\begin{tabular}{lllll}
  \hline 
  \noalign{\smallskip} 
  \multicolumn{1}{c}{Line} & \multicolumn{1}{c}{Wavelength}
  & \multicolumn{1}{c}{$E_{\rm up}/k^{a}$}
  & \multicolumn{1}{c}{$n_{\rm crit}^{b}$}
  & \multicolumn{1}{c}{$A^{c}$}\\
  & \multicolumn{1}{c}{($\mu m$)}
  & \multicolumn{1}{c}{(K)}
  & \multicolumn{1}{c}{(cm$^{-3}$)}
  & \multicolumn{1}{c}{(s$^{-1}$)}\\
  \noalign{\smallskip} 
  \hline
  \noalign{\smallskip} 
  O{\sc I}\ $^3$P$_1 \rightarrow ^3$P$_2$ & 63.183 & 227.72& 4.7$\times 10^{5}$  & 8.87$\times 10^{-5}$ \\
  O{\sc I}\ $^3$P$_0 \rightarrow ^3$P$_1$ & 145.525 & 326  & 1.0$\times 10^{5}$ &  1.74$\times 10^{-5}$ \\
  C{\sc II}\ $^2$P$_{3/2} \rightarrow ^2$P$_{1/2}$ & 157.740 & 91.22 & 2.8$\times 10^{3}$ & 2.4$\times 10^{-6}$ \\
  \noalign{\smallskip} 
  \hline
\end{tabular}
\begin{flushleft}
  $^{a}${The energy of the upper state of the transition
    relative to the ground state in temperature.} \\
  $^{b}${The critical density is $A/\gamma$, where $A$ is the
    Einstein-A coefficient and $\gamma$ the collision rate
    coefficient. The critical densities are computed assuming LTE,
    $T_{\rm kin}$=100~K and the optically thin limit.}\\
  $^{c}$ Einstein-A coefficients were taken from \cite{Galavis97}.\\
\end{flushleft}
\end{table}

\begin{figure*}[!ht]  
\includegraphics[scale=0.38,angle=90]{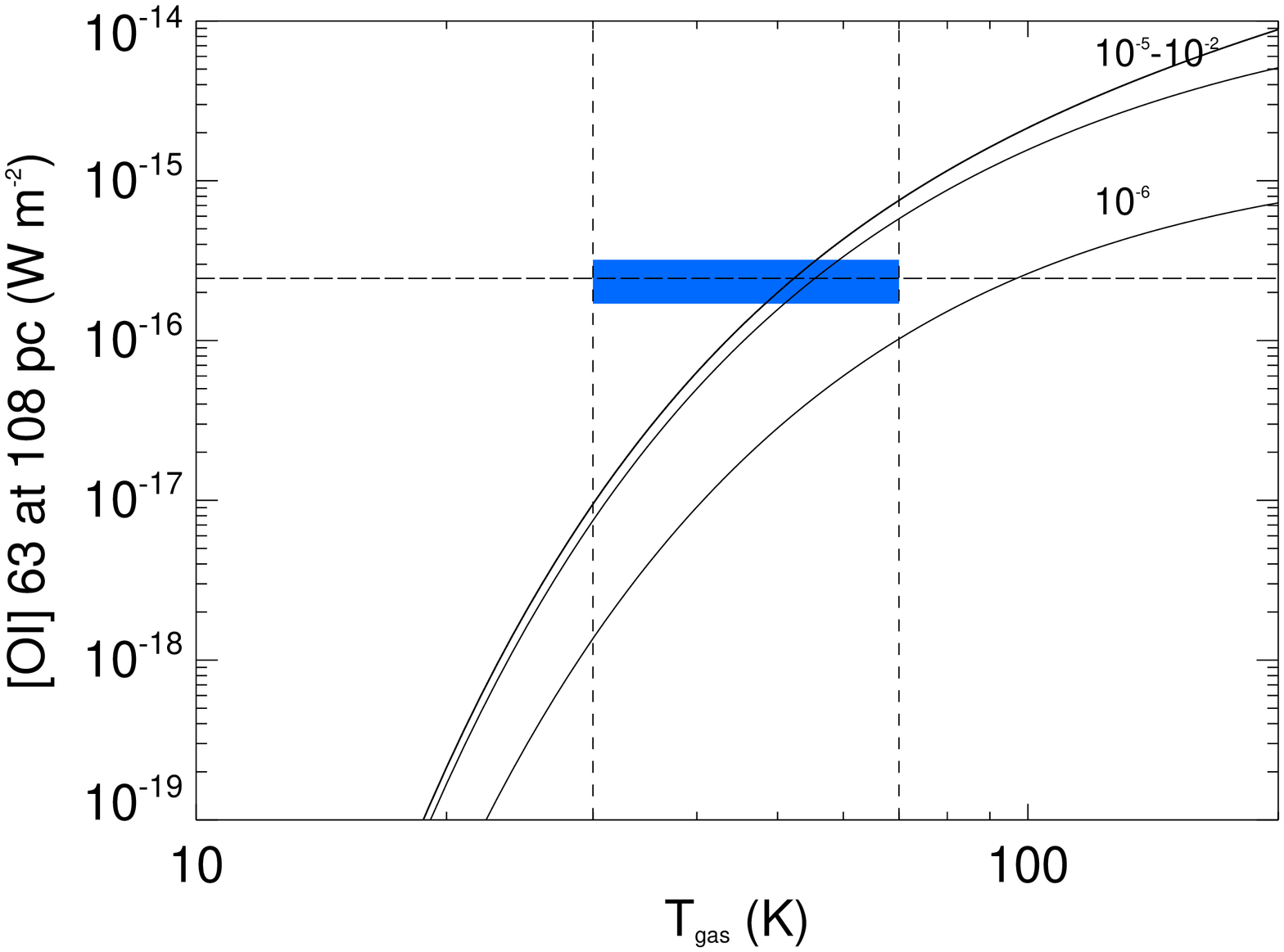}
\includegraphics[scale=0.38,angle=90]{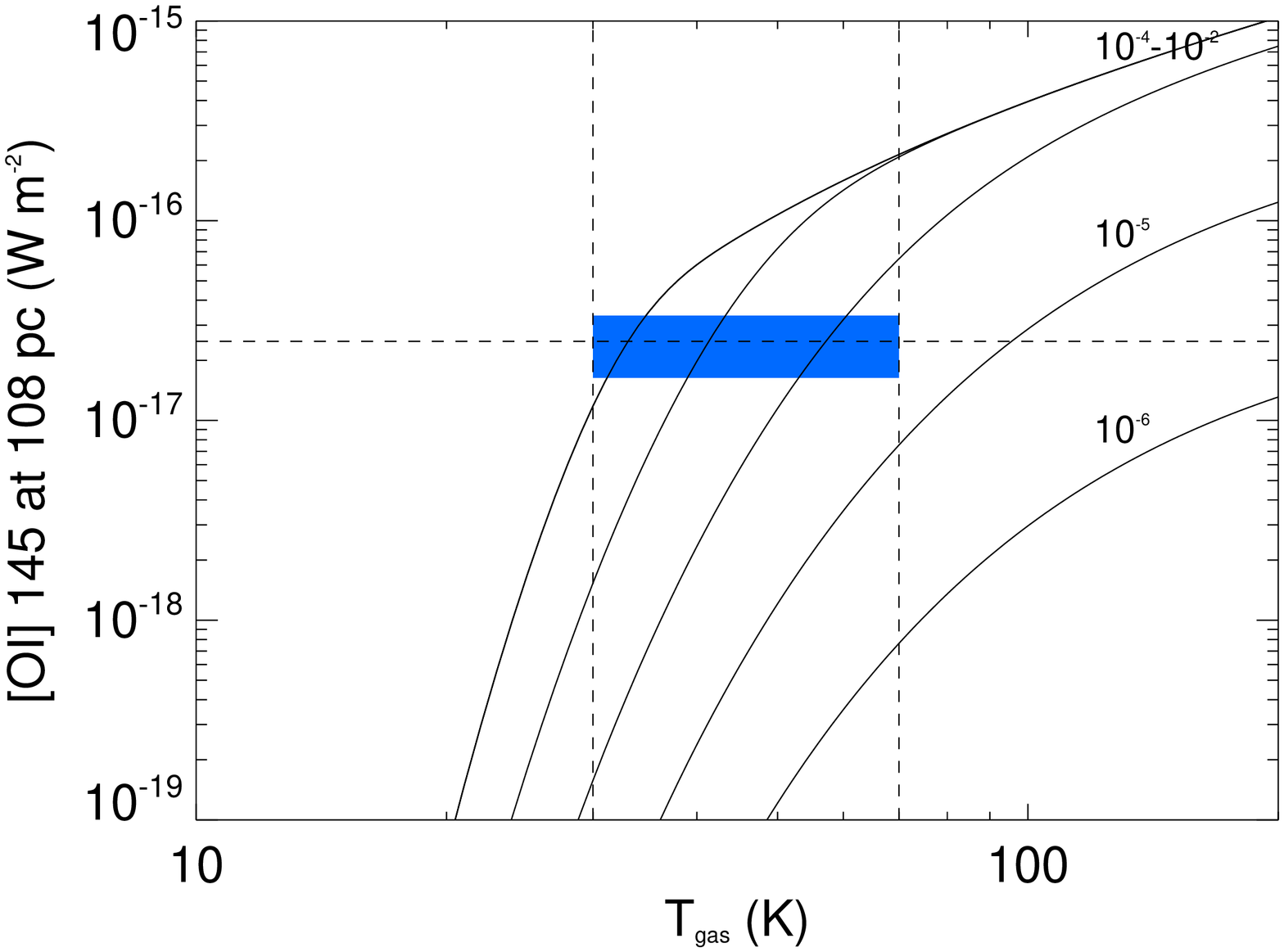}
\includegraphics[scale=0.38,angle=90]{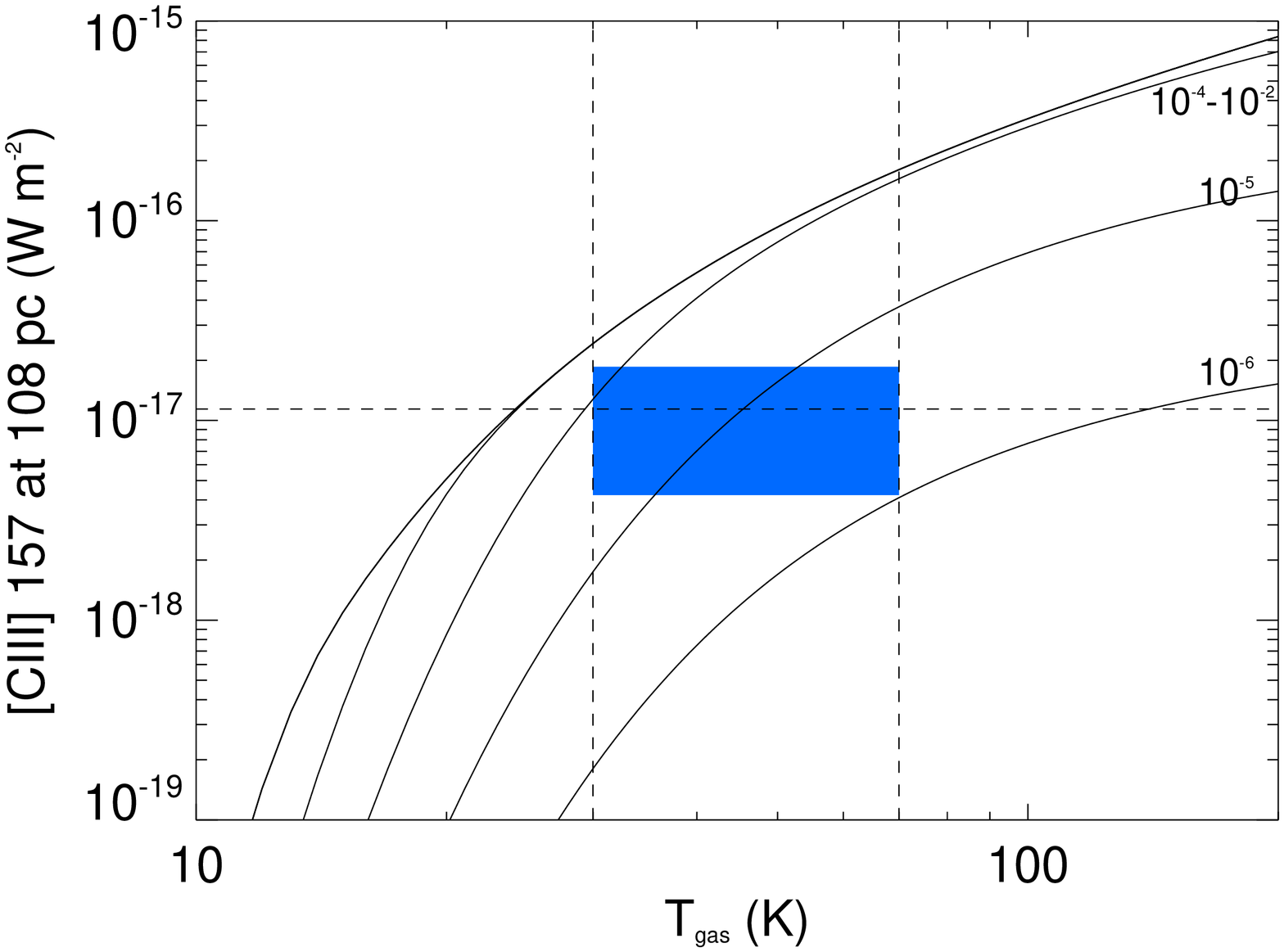}
\includegraphics[scale=0.38,angle=90]{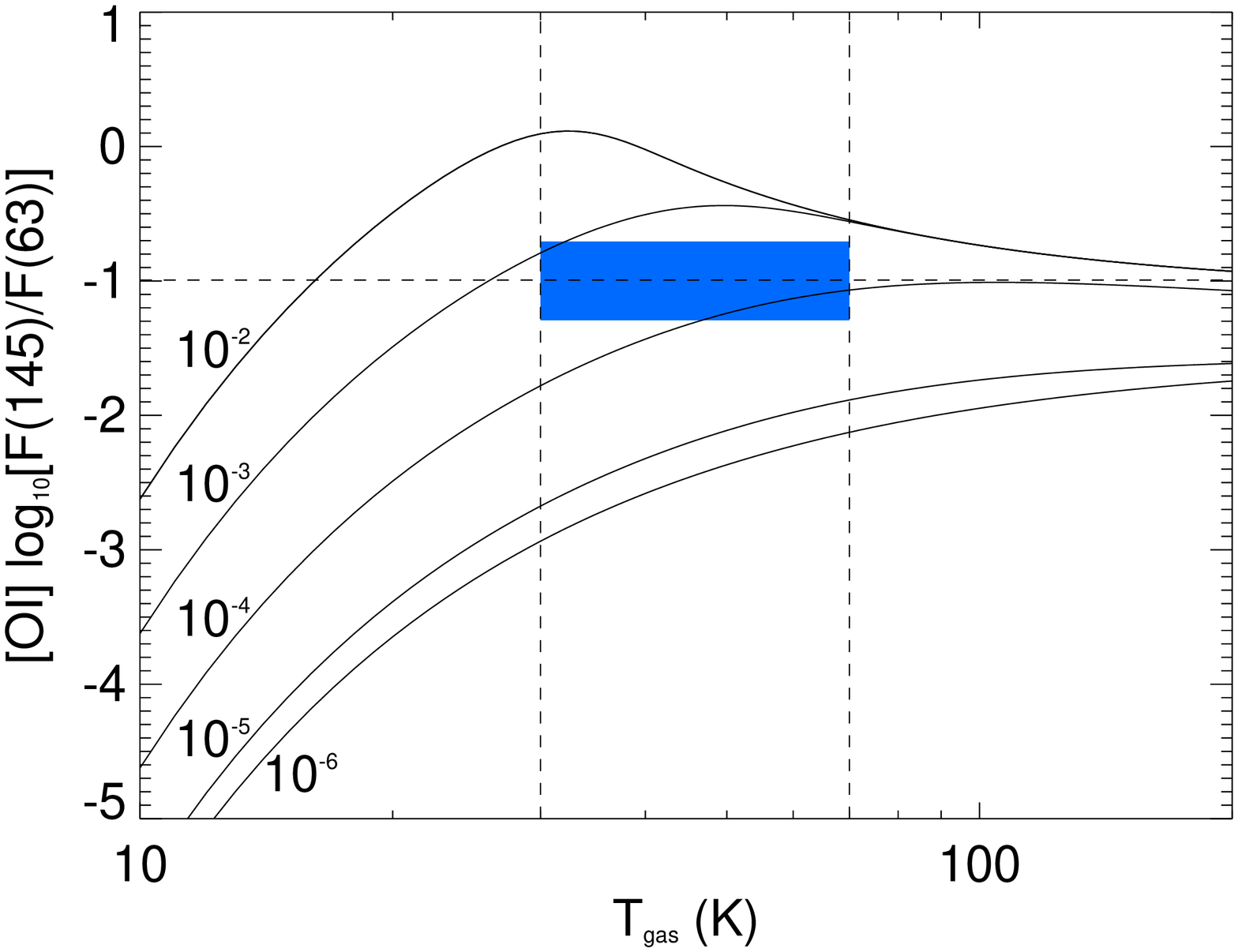}
\caption{Predicted [\OI]\ 63 and 145 $\mu$m and [\CII]\ 157 $\mu$m
  fine-structure fluxes as a function of the gas temperature. The
  fourth panel shows the predicted ratio between the [\OI]\ 63 and 145
  micron fluxes. The observed ratio is overplotted by the horizontal
  dashed lines. The total disk gas masses in M$_\odot$ are
  indicated. The blue box encompasses gas temperatures between 30 and
  70~K and the observed [\OI]\ at 63$\mu$m, [\OI]\ at 145$\mu$m,
  [\CII]\ at 157$\mu$m, and the $F$([\OI]145$\mu$m)/$F$([\OI]63$\mu$m)
  flux with 3~$\sigma$ error bars.}
  \label{fig_simple_model}                  
\end{figure*}    
The oxygen fine-structure line ratio 63 $\mu$m/145 $\mu$ is $\sim$9.8.
A ratio of about 10 has been found for a large number of young stellar
objects \citep{Liseau2006A&A...446..561L}. A ratio of $\sim$~10 can be
reproduced theoretically by optically thick
emissions. \citet{Liseau2006A&A...446..561L} proposed that the 63
$\mu$m line can be re-absorbed by intervening cloud material.  The two
atomic oxygen and the ionized carbon fine-structure lines were
detected. At first we were able to constrain the gas mass and average
gas temperature from the line fluxes and their ratios. We modeled the
[\OI]\ and [\CII]\ emissions with a simple model where the disk has a
constant surface density and a radius $R_{\mathrm{out}}$. Given the
disk mass and elemental abundances for atomic oxygen O and C$^+$
\citep{Woitke2009A&A...501..383W}, the vertical optical depth $\tau$
can be computed. We assumed that atomic oxygen and singly ionized
carbon are the most abundant species that contain oxygen and carbon,
respectively. This assumption is only valid for low-mass disks.  The
atomic oxygen abundance is $\chi$(O)=8.5$\times$10$^{-4}$ and the
ionized carbon abundance is $\chi$(C)=3.5$\times$10$^{-4}$. In an
oxygen-rich environment where the gas density is low, chemical models
show that atomic oxygen is by far the main carrier of oxygen. The
assumed C$^{+}$ abundance is an upper limit to the actual value since
carbon can occur in the form of atomic carbon and CO. The line fluxes
in (W m$^{-2}$) are computed using the equation
\begin{equation}
\nu F_\nu \simeq B_\nu(T)(1-e^{-\tau})\frac{\pi R_{\mathrm{out}}^2}{d^2} \Delta\nu,
\end{equation}  
where $B_\nu(T)$ is the Planck function (W m$^{-2}$ Hz$^{-1}$) at a
single average temperature $T$, which is a parameter of the model,
$\tau$ is the optical depth that depends on the inclination of the
disk,
\begin{equation}
  \tau(\mathrm{O,C^+}) = \frac{c^2}{8\pi}\frac{A}{\nu^2}\left(\frac{1}{\Delta \nu}\right)\left(e^{\left(h\nu/kT\right)}-1\right)\frac{N(\mathrm{O,C^+})}{\cos{i}}, 
\end{equation}
$R_{\mathrm{out}}$ is the outer radius, $d$ is the distance
($R_{\mathrm{out}}$ and $d$ have the same units of length), and
$\Delta
\nu=\nu/c\sqrt{\mathrm{v}_{\mathrm{therm}}^2+\mathrm{v}_{\mathrm{turb}}^2}$
is the line width, with $\nu$ the frequency of the transition in
Hz. $N(\mathrm{O,C^+})$ is the oxygen or ionized carbon (C$^{+}$)
column density.

We adopted an outer radius of 500~AU and a turbulent width
v$_{\mathrm{turb}}$ of 0.15 km s$^{-1}$. A value of 0.3 km s$^{-1}$
has been found in the disk around \object{HD163296}
\citep{Hughes2011ApJ...727...85H}.

The values of the Einstein coefficients $A$ and other molecular
parameters are given in Table~\ref{line_parameters}. The inclination
is 55\degree. The disk mass is
\begin{equation}
M_{\mathrm{gas}}=\pi R_{\mathrm{out}}^2 \mu_{\mathrm{gas}} \frac{N(\mathrm{O,C^+})}{\chi(\mathrm{O,C^+})},
\end{equation}
where $\mu_{\mathrm{gas}}$= 2.2 $\times$ the atomic mass is the mean
molecular mass.

\begin{figure}[!ht]  
\resizebox{\hsize}{!}{\includegraphics[angle=90]{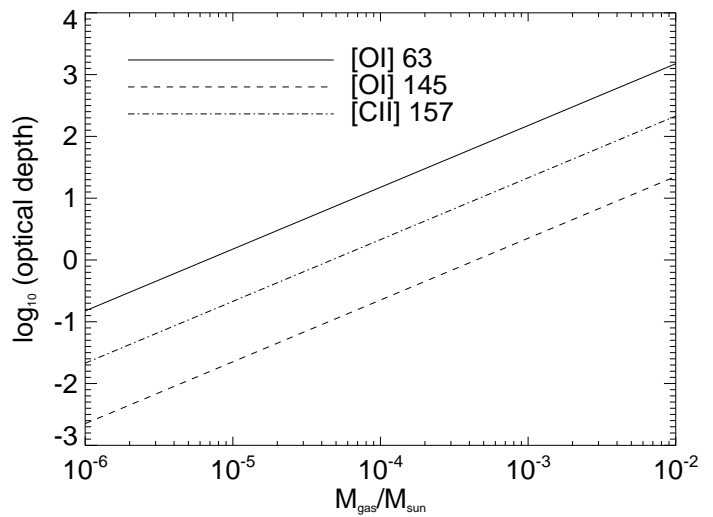}}
\caption{[\OI]\ 63 and 145 $\mu$m and [\CII]\ 157 $\mu$m line optical
  depths as a function of the gas disk mass at
  $T_{\mathrm{gas}}$=~60~K.}
  \label{fig_optical_depths}                  
\end{figure}

We compare the line flux and line ratio predictions with the observed
values and ratios in Fig.~\ref{fig_simple_model}. We can bracket the
average disk gas mass between 1$\times$10$^{-4}$ and
5$\times$10$^{-4}$ M$_\odot$ and the gas temperature between 30 and
70~K. The line optical depths as a function of the disk gas mass for a
gas temperature of 60~K are shown in
Fig.~\ref{fig_optical_depths}. The [\OI]\ 63 micron emission is
optically thick for disks with gas masses greater than 10$^{-5}$
M$_\odot$, the [\CII]\ emission becomes optically thick for gas masses
greater than 10$^{-4}$ M$_\odot$, and the [\OI]\ 145 micron emission
for gas masses greater than $\sim$~5$\times$10$^{-4}$ M$_\odot$.

\begin{figure*}[!ht]  
\includegraphics[scale=0.38,angle=90]{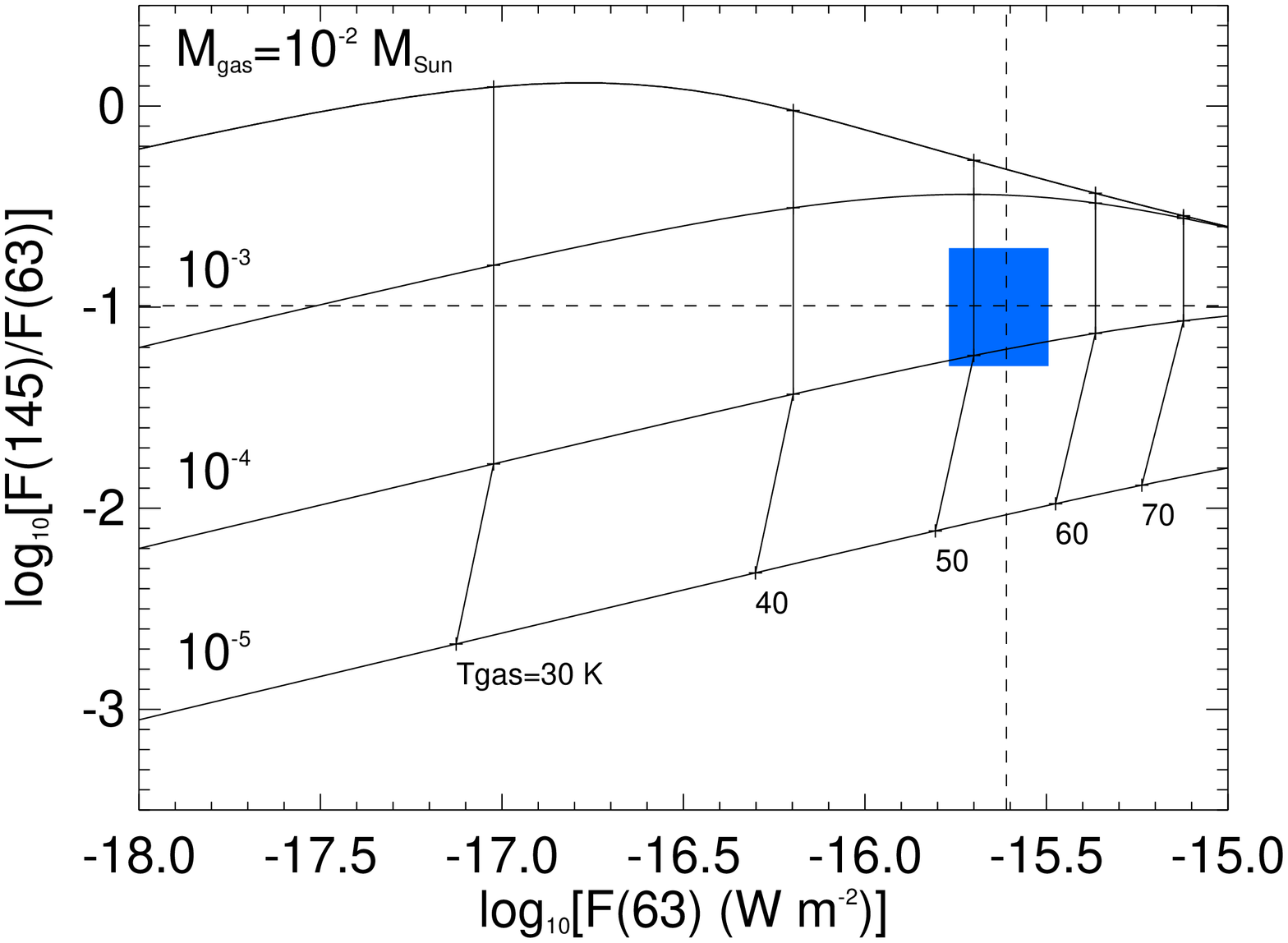}
\includegraphics[scale=0.38,angle=90]{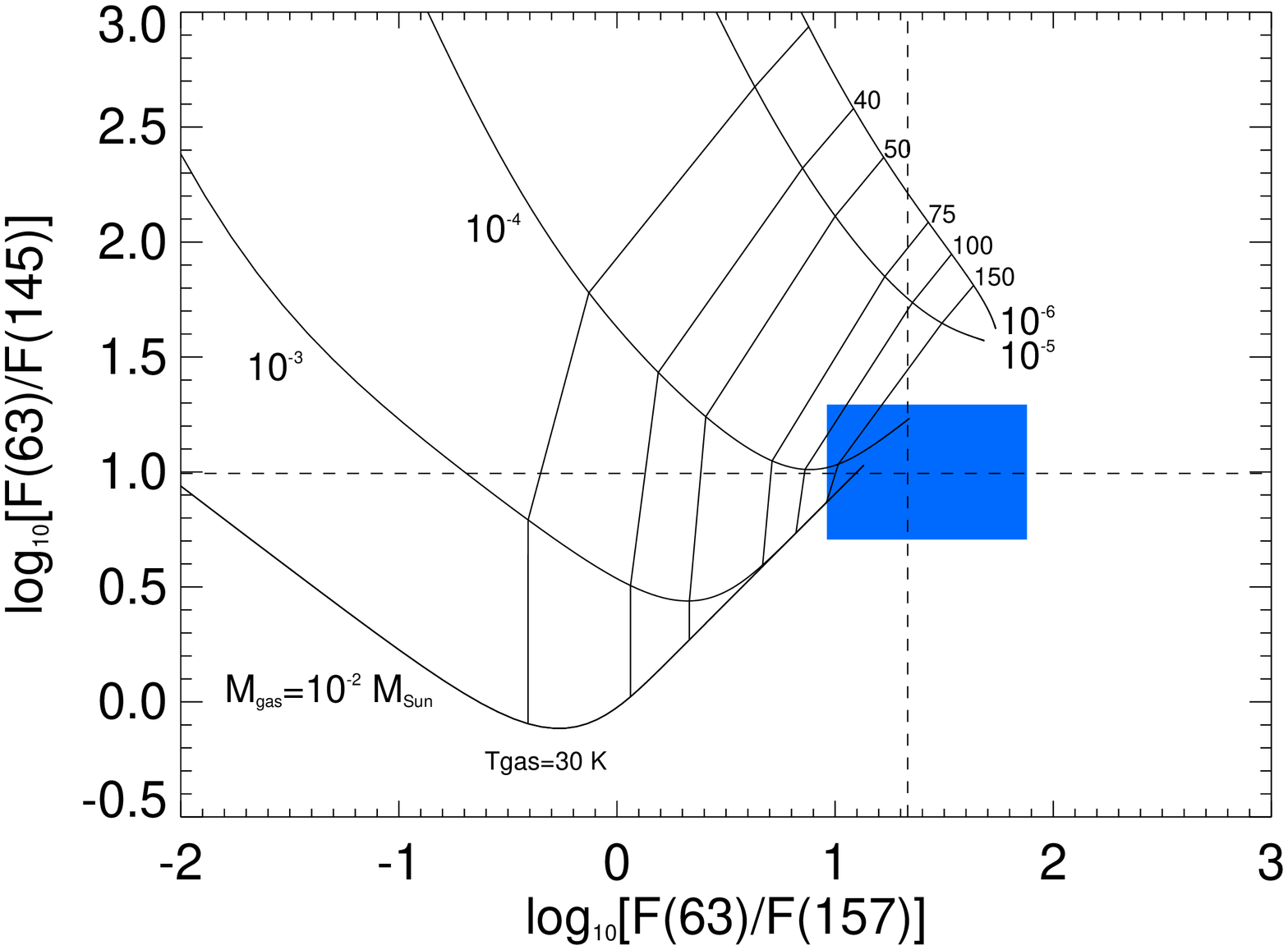}
\caption{Oxygen fine-structure line flux ratio as a function of
  [\OI]\ 63 microns is shown in the left panel. The right panel shows
  the [\OI]\ 63/[\OI]\ 145 ratio versus the [\OI]\ 63/[\CII]\ 157
  ratio. The blue boxes indicate the observed flux ratios. All
  ratios suggest a gas mass of a few 10$^{-4}$ M$_\odot$ and a
  disk-averaged temperature of 75-150~K.}
  \label{fig_simple_model2}                  
\end{figure*}  
\section{Disk continuum-emission
  modeling}~\label{continuum_modeling}

The aim of the paper is not to provide a perfect fit to the continuum
data, nor to uniquely constrain the dust grain composition, but to
constrain the gas- and dust (solid) disk masses. In particular, we did
not attempt to fit the PAH features in detail. The continuum
radiative-transfer computes the mean intensity inside the disk, which
is essential for accurately characterizing the photoelectric and
photoreaction processes.  In addition, the fit to the SED constrains
the grain surface area and temperature needed for the grain-surface
interactions and adsorption/desorption processes.

\subsection{Stellar properties}\label{stellar_properties}

Accurate knowledge of the stellar properties is paramount for
modeling the SED and gas lines. \source\ is a B9.5V star with an
effective temperature of 10,000~K \citep{Merin2004A&A...419..301M}
located at a distance of 99~pc
\citep{vandenancker1997A&A...324L..33V}. Alternatively,
\citet{Merin2004A&A...419..301M} located \source\ at 108~pc. The
stellar parameters are summarized in the upper part of
Table~\ref{disk_parameters}.

Broad ($FWHM=$~154~km s$^{-1}$) optical \OIfs emission at 6300 \AA\
centered at the stellar velocity has been detected
\citep{Acke2005A&A...436..209A} with 2.9$\times$10$^{-4}$ L$_\odot$,
which translates into 9.5$\times$10$^{-16}$ W m$^{-2}$. In addition, the
H$\alpha$ emission has an equivalent width of $EW$=-6.7 \AA.\ Both
line profiles are double-peaked.

The star is not variable, either spectroscopically in the optical
\citep{Mendigutia2011A&A...535A..99M}, or photometrically in the
mid-IR \citep{Kospal2012ApJS..201...11K}. Therefore it is relatively
safe to use non-simultaneous photometric data. The non-variability of
the star supports the idea that \source\ is more evolved than typical
HerbigAe stars.

\subsection{SED modeling}\label{sed_modeling}

\subsubsection{Photometric data}\label{photometric_data}

We augmented the {\it Herschel} photometric and spectroscopic
continuum data with measurements taken from the literature, which are
summarized in Table~\ref{table_photometry}. These data were taken with
various telescopes and beam sizes. The photometric fluxes taken with
large-beam telescopes are systematically higher than the fluxes
obtained with small-beam telescopes. This trend suggests a
non-negligible flux contribution from foreground and background
emissions. Therefore we chose to use data points from small-beam
observations wherever possible. 

We fitted the spectral energy distribution (SED) with the Monte Carlo
continuum radiative transfer code {\sc MCFOST}
\citep{Pinte2006A&A...459..797P,Pinte2009A&A...498..967P}. 

The mid-IR spectrum of \source\ is dominated by prominent PAH features
with a lack of the silicate emission feature at $\lambda
\simeq$~10~$\mu$m
\citep{Li2003ApJ...594..987L,Sloan2005ApJ...632..956S,Acke2010ApJ...718..558A}. In
the absence of unambiguous constraints, we did not try to find the
exact dust composition in the disk. Instead, we assumed pure amorphous
magnesium-rich olivine (Mg$_2$SiO$_4$) grains
\citep{Jager2003A&A...408..193J}. We did not include carbonaceous
material such as amorphous carbon or graphite, nor did we include
water ice. The dust size-distribution is a power-law defined by a
minimum radius $a_{\mathrm{min}}$, maximum radius $a_{\mathrm{max}}$,
and power-law index $p$. The lack of the prominent 10-$\mu$m silicate
feature hints at the presence of grains larger than a few
micrometres. Furthermore to match the 1~mm flux, the maximum grain
radius is probably a least a few millimeters. We adopted values for
$a_{\mathrm{max}}$ between 0.5 and 1 cm. There are most likely not
many solid objects larger than one centimeter in this
collision-dominated disk, where the dust collision timescale is
shorter than the transport timescale
\citep{Muller2010ApJ...708.1728M}. The final values for the minimum
and maximum grain radii are given in Table~\ref{disk_parameters}.

\begin{figure}[!ht]
\centering
\resizebox{\hsize}{!}{\includegraphics[]{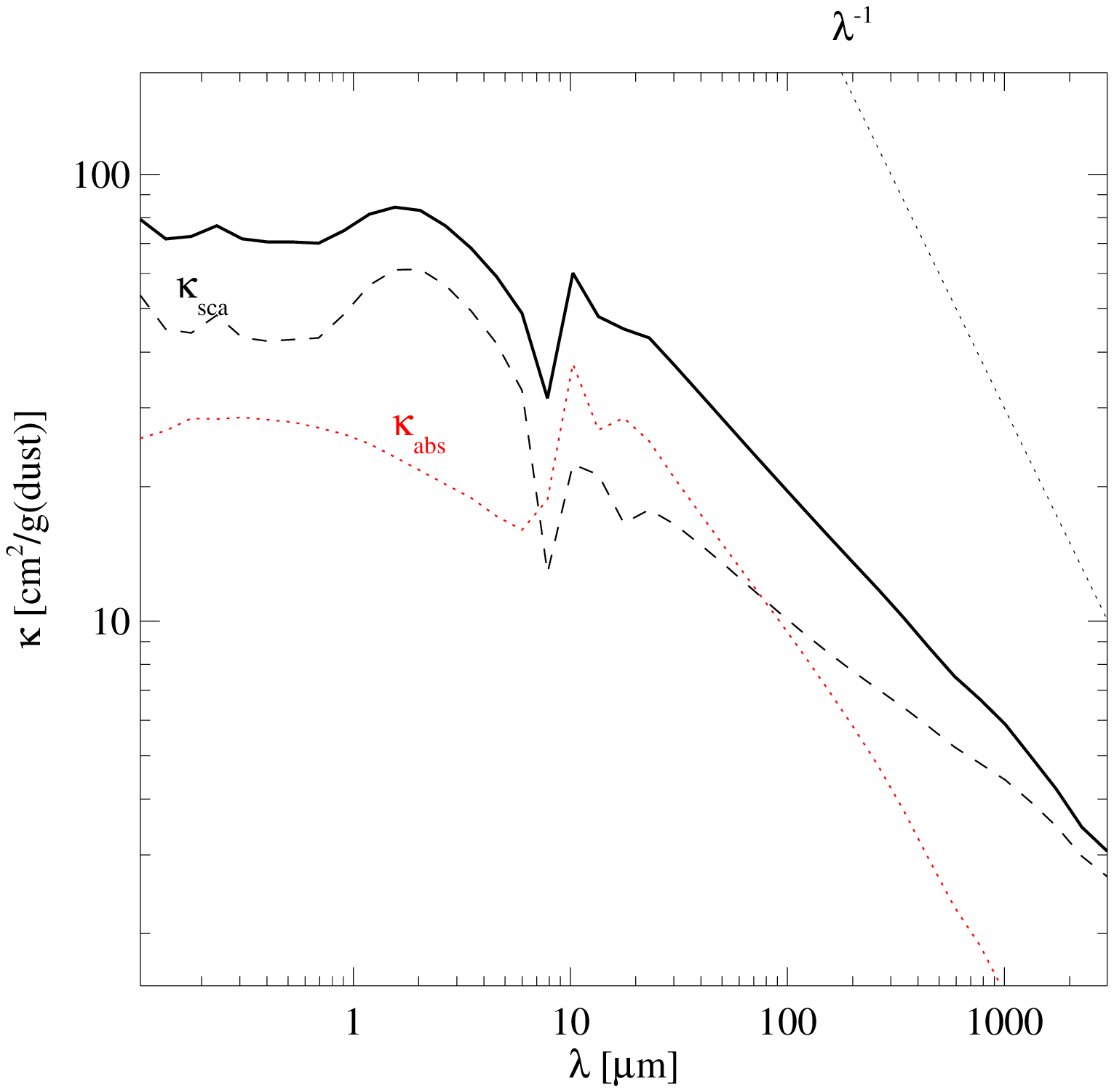}}
\caption{Dust opacity. The dust grains are made of astronomical
  silicates.}
\label{fig_kappa}          
\end{figure}

The dust opacity was computed according to the Mie theory for compact
spherical grains and is shown in Fig.~\ref{fig_kappa} for the size
distribution parameters in Table~\ref{disk_parameters}. We used a
silicate mass density of 3.5 g cm$^{-3}$. The extinction is dominated
by dust scattering from the UV to $\lambda \simeq$~10~$\mu$m. We
distinguished the dust grain mass, which includes the mass of grains
up to 1 mm in radius, and the total solid mass, which corresponds to
the mass of all the solids.

\subsubsection{Modeling of the PAH features and image}\label{PAH_modeling}

We used circumcircumcoronene cation (C$_{150}$H$_{30}^+$,
size$\sim$6.85 \AA) as our typical PAH, because only large PAHs can
survive in HerbigAe disks \citep{Visser2007A&A...466..229V}.
\citet{Stein85} found that large compact PAHs such as
circumcircumcoronenes are extremely stable at the high temperatures
(1000~K-3000~K) reached after the absorption of a UV photon. The
circumcircumcoronenes can be singly negatively charged, neutral, and
up to three times positively charged. In addition, the relative
strengths between the PAH features in \source\ suggest that the PAHs
are large and mostly positively charged
\citep{Sloan2005ApJ...632..956S, Bauschlicher2008ApJ...678..316B,
  Bauschlicher2009ApJ...697..311B}. The PAH excitation and emission
mechanisms implemented in the radiative transfer code {\sc MCFOST}
follow the opacity model of \citealt[][]{Draine2001ApJ...551..807D},
\citealt[][]{Draine2001ApJ...554..778L}, and
\citealt[][]{Draine2007ApJ...657..810D}. We modeled the PAH emission
profile at 8.6 $\mu$m for comparison with the observed profile.

\begin{figure*}[!ht]
\centering
\includegraphics[scale=0.5,angle=0]{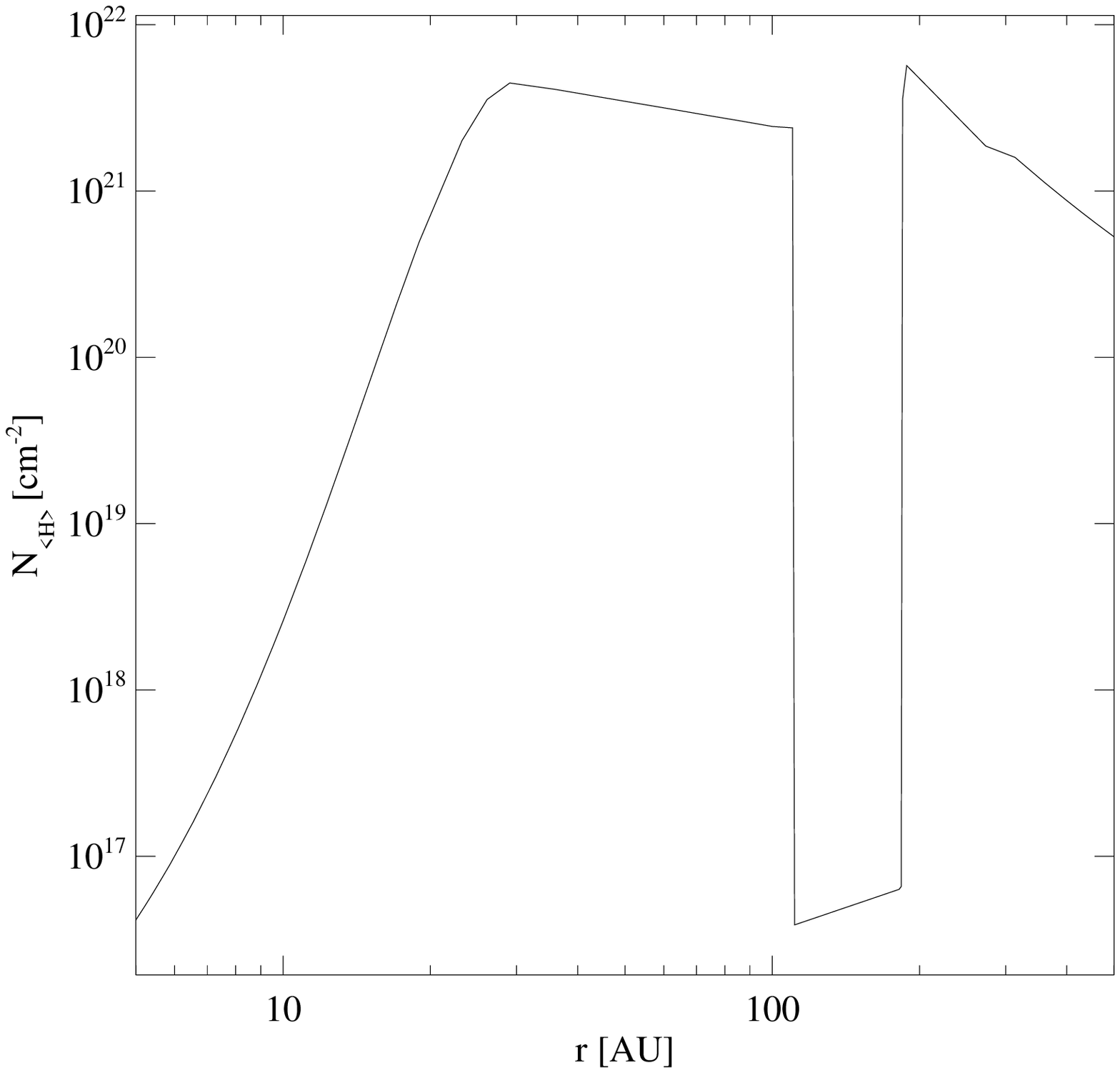}
\includegraphics[scale=0.5,angle=0]{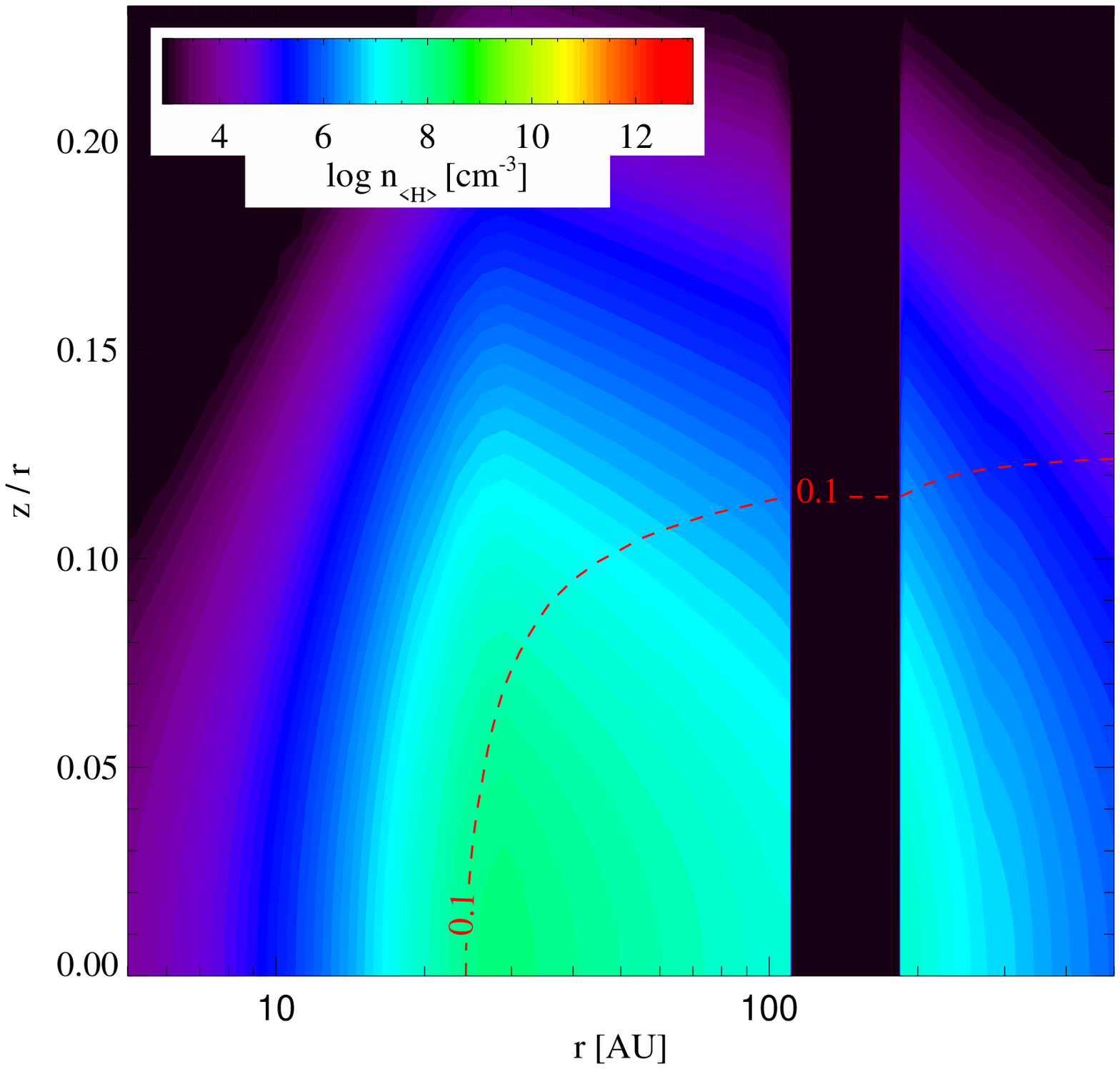}
\caption{Adopted half-disk surface density profile (left panel) and
  gas density structure for a $H_0$=5\% opening and a gas-to-dust-mass
  ratio of 100 disk, ie $M_{\mathrm{gas}}$=4.9$\times$ 10$^{-4}$
  M$_\odot$ (right panel). The gap between 110 and 185~AU is not
  entirely devoid of gas and dust.}
\label{fig_surf_density_profile}          
\end{figure*}  
\begin{figure}[!ht]
\centering
\resizebox{\hsize}{!}{\includegraphics[]{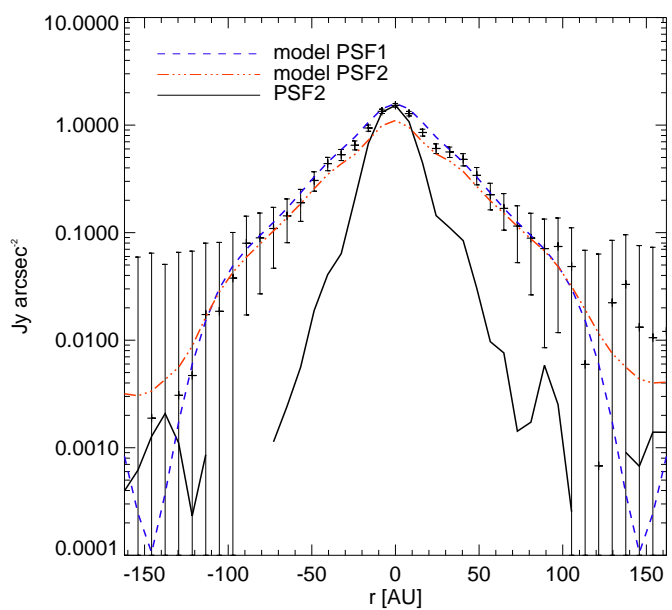}}
\caption{Fit to the PAH radial emission profile at 8.6 $\mu$m along
  the major axis obtained with VISIR at the VLT. The same model has
  been convolved with the observed or theoretical noise-free PSF. The
  theoretical PSF fits the core of the observed function. The model
  succeeded in fitting the profile.}
\label{fig_PAH_profile}          
\end{figure}  
\subsubsection{Disk structure}\label{disk_structure}

We adopted a parametric disk for the disk geometry. We assumed a
modified version of the surface density profile of
\citet{Li2003ApJ...598L..51L} constrained by imaging studies. 

\begin{figure*}   
\sidecaption
\includegraphics[angle=90,width=12cm]{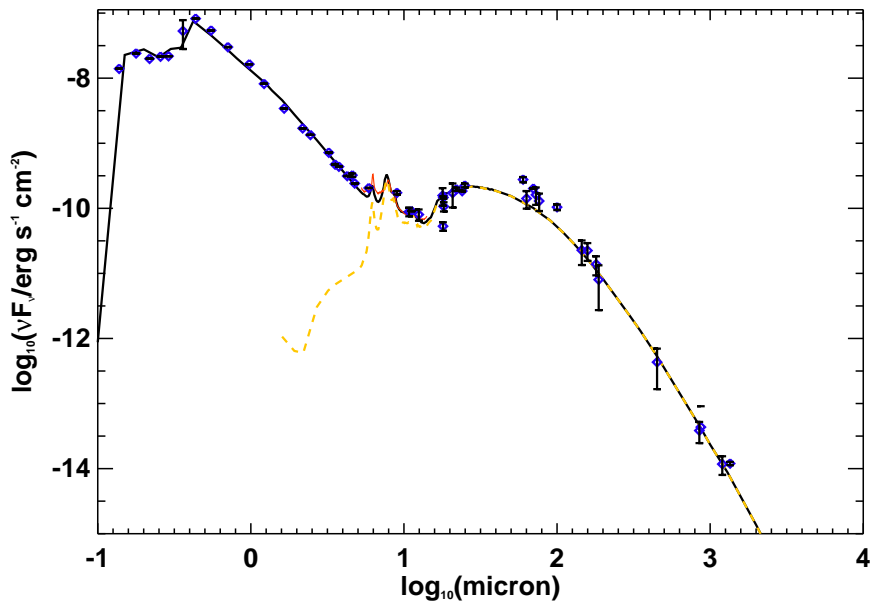}
\caption{Fit to the SED and the {\it Spitzer-IRS} spectrum by a disk
  model with $H_0$=5\%. The photometric points are shown with
  3$\sigma$ error bars. }
  \label{fig_SED_ProDiMo_fit}  
\end{figure*}

The best-fit SED is shown in Fig.~\ref{fig_SED_ProDiMo_fit}.  The disk
around \source\ is composed of three major rings peaking at $\sim$~15,
185, and 300 AU, respectively (see Table~\ref{disk_parameters}).  The
disk surface density and gas density structure of one of the disk
models are show in Fig.~\ref{fig_surf_density_profile}. In that
  figure, the two outer rings merge into the outer disk. The location
of the rings matches the scattered-light images. The disk outer radius
($R_{\mathrm{out}}$) is located at 500~AU, consistent with
scattered-light images in the optical
\citep{Clampin2003AJ....126..385C}.  The gap between 100 and 185~AU is
consistent with near-IR scattered-light and 8.6~$\mu$m images. The
disk's surface geometry is defined by an opening angle $H_0$ at a
given radius $R_{\mathrm{ref}}$ and a flaring index $\gamma$=1 (i.e.,
no flaring) such that the gas scale-height is given by
$H=H_0(R/R_{\mathrm{ref}})^{\gamma}$. The scale-height of optically
thin disks cannot be well constrained by the fit to the SED and is
thus a free parameter for both the continuum and gas
modeling. Therefore we ran a series of models with varying
scale-height $H_0$ = 3, 5, 10, and 20~AU at the reference radius
$R_{\mathrm{ref}}$= 100~AU and assumed that the gas and dust are
well-mixed. The input {\sc Phoenix} stellar spectrum plotted in red is
taken from \citet{Brott2005ESASP.576..565B}. The disk inclination $i$
at 55\degree\ with respect to the rotation axis is well constrained by
imaging data
\citep{Augereau1999A&A...350L..51A}. \citet{Weinberger2002ApJ...566..409W}
independently found an inclination of 51\degree. The UV spectrum of
\source\ can be entirely explained by the stellar emission without
excess from gas accretion. Therefore the UV excess parameter, which
accounts for non-photospheric UV emission, was set to 0.

\begin{table*}
\begin{center}
  \caption{Disk modeling parameters.}\label{disk_parameters}
		\begin{tabular}{lllll}
                  \hline
                  \noalign{\smallskip}   
                  & \multicolumn{2}{c}{Fixed parameters} & &   \\
                  \noalign{\smallskip}   
                  Distance     & $d$ (pc)              &  108 & &\\
                  Spectral type &   & B9.5V &&\\
                  Stellar mass & $M_*$ (M$_\odot$)       &  2.0 & &\\
                  Stellar luminosity &$L_*$ (L$_\odot$)  &  25.77 &&\\ 
                  Effective temperature & $T_{\mathrm {eff}}$ (K) & 10000&&\\
                  Age & Myrs & 4.71 \\
                  Solid material mass density & $\rho_{\mathrm{dust}}$ (g cm$^{-3}$) & 3.5&& \\
                  ISM UV field   & $\chi$          &  1.0&&\\
                  $\alpha$ viscosity parameter      & $\alpha$           & 0.0&&\\
                  Non-thermal velocity & $v_{\mathrm{turb}}$ (km s$^{-1}$) & 0.15&&\\
                  UV excess                         & $F_{\mathrm{UV}}$ & 0.0&&\\
                  Cosmic ray flux                  & $\zeta$ (s$^{-1}$)           & 1.7 $\times$ 10$^{-17}$&&\\
                  Disk inclination & $i$ (\degr) & 55&&\\
                  Extinction       & $E$(B-V) & 0.095&&\\
                  & $R_{\mathrm{V}}$ & 3.1& &  \\
                  \noalign{\smallskip}   
                  \hline
                  \noalign{\smallskip}   
                    & &  Inner disk & Outer disk          & Outer disk\\
                    & &       disk  & 1$^{\mathrm{st}}$ ring & 2$^{\mathrm{nd}}$ ring\\ 
                  \noalign{\smallskip}   
                  \cline{3-5}\\  
                  & & \multicolumn{3}{c}{{\sc MCFOST} parameters}\\
                  \noalign{\smallskip}
                  Inner radius                      & $R_{\mathrm{in}}$ (AU)  &  5   & 185  & 300 \\
                  Outer radius                      & $R_{\mathrm{out}}$ (AU) & 110  & 500 & 500  \\
                  Column density index  & $\epsilon$ &   1 & 3 & 1 \\
                  Reference scale height            & $H_0$ (AU)           & \multicolumn{3}{c}{3, 5, 10, 20}\\
                  Reference radius                  & $R_{\mathrm{ref}}$ (AU)                & \multicolumn{3}{c}{100}\\
                  Flaring index                     & $\gamma$            & \multicolumn{3}{c}{1}\\
                  Minimum grain size       & $a_{\mathrm{min}}$ ($\mu$m) & \multicolumn{3}{c}{0.5}\\
                  Maximum grain size       & $a_{\mathrm{max}}$ (cm) & 1 & 0.5 & 0.5\\
                  Dust size distribution index  & $p$               & \multicolumn{3}{c}{3.5}\\
                  Dust mass ($a\le$1 mm) & $M_{\mathrm{dust}}$ (M$_{\odot}$)                   & 6.2 $\times$ 10$^{-8}$ & 1.8 $\times$ 10$^{-6}$ & 3.1 $\times$ 10$^{-7}$ \\
                  Solid mass           &$M_{\mathrm{solid}}$ (M$_{\odot}$)                   & 2.0 $\times$ 10$^{-7}$  & 4.0 $\times$ 10$^{-6}$ & 7.0 $\times$ 10$^{-7}$\\ 
                  PAH  mass          &$M_{\mathrm{PAH}}$(M$_{\odot}$)             & 2.0 $\times$ 10$^{-11}$ & 1.2 $\times$ 10$^{-10}$ & 2.1 $\times$ 10$^{-11}$ \\
                  \noalign{\smallskip}   
                  \cline{3-5}\\  
                  \noalign{\smallskip}   
                  & & \multicolumn{3}{c}{{\sc ProDiMo} parameters}\\
                  \noalign{\smallskip}   
                  Disk gas mass                     & $M_{gas}$ (M$_{\odot}$) &  \multicolumn{3}{c}{4.9 $\times$ 10$^{-5}$ -- 4.9 $\times$ 10$^{-4}$}\\
                  \noalign{\smallskip}   
                  \hline
\end{tabular}
\ \\
The distance and stellar parameters are taken from \citep{Merin2004A&A...419..301M}. 
\end{center}
\end{table*}

\begin{table}
\begin{center}
  \caption{PAH abundances $f_{\mathrm{PAH}}$ relative to the ISM
    abundance of 3 $\times$ 10$^{-7}$.}\label{PAH_abundance}
		\begin{tabular}{llll}
                  \toprule
                  & inner disk & outer disk\\
\noalign{\smallskip}   
\hline
                  \noalign{\smallskip}   
                  $M_{\mathrm{solid}}$ (M$_{\odot}$)  & 2.0$\times$10$^{-6}$ & 4.7$\times$10$^{-6}$\\
                  $M_{\mathrm{PAH}}$ (M$_{\odot}$)    & 2.0$\times$10$^{-11}$ & 1.4$\times$10$^{-10}$\\
                  PAH                           & \multicolumn{2}{c}{C$_{150}$H$_{30}$}\\
                  $a_{\mathrm{PAH}}$ (\AA)          & \multicolumn{2}{c}{6.84}  \\
                  $m_{\mathrm{PAH}}$ (a.m.u.)       & \multicolumn{2}{c}{1830}   \\
                  \noalign{\smallskip}   
                  \hline
                  \noalign{\smallskip}   
                  $M_{\mathrm{gas}}/M_{\mathrm{solid}}$ & \multicolumn{2}{c}{$f_{\mathrm{PAH}}$}\\

\noalign{\smallskip}   
\hline
\noalign{\smallskip}   
                   100    &   2.0 $\times$ 10$^{-3}$ & 6.7 $\times$ 10$^{-3}$ \\
                    50    &   3.9 $\times$ 10$^{-3}$ & 1.3 $\times$ 10$^{-2}$ \\
                    20    &   9.9 $\times$ 10$^{-3}$ & 3.3 $\times$ 10$^{-2}$ \\
                    10    &   2.0 $\times$ 10$^{-2}$ & 6.7 $\times$ 10$^{-2}$\\   
\noalign{\smallskip}     
\bottomrule
\end{tabular} 
\end{center}
\end{table}

The best fit to the SED is shown in Fig.~\ref{fig_SED_ProDiMo_fit} for
the parameters listed in Table~\ref{disk_parameters} and
$H_0$~=~5\,AU.  The adopted surface density profile provides good fits
to the SED and an 8.6~$\mu$m radial profile. The synthetic radial
emission profile at 8.6 $\mu$m is compared with the radial profile
observed with VISIR at the VLT (Fig.~\ref{fig_PAH_profile}).

The inferred dust mass in grains with radius $\leq$~1\,mm is
$M_{\mathrm{dust}}\simeq$~2.1 $\times$~10$^{-7}$\, M$_{\odot}$ and the
total mass in solids (from nanograins to pebbles) up to
$a_{\mathrm{max}}$~=~1\,cm is $M_{\mathrm{solid}}\simeq$~4.9 $\times$
10$^{-6}$\,M$_{\odot}$, similar to the values found by
\cite{Li2003ApJ...594..987L} ($\simeq$~6.7 $\times$
10$^{-6}$\,M$_{\odot}$) and \cite{Merin2004A&A...419..301M}
($\simeq$~6.4 $\times$ 10$^{-6}$\,M$_{\odot}$). The total mass of
PAHs, assuming a unique size of $a_{\mathrm{PAH}}$~=~6.85\,\AA, is
$\simeq$~1.6 $\times$ 10$^{-10}$\,
M$_{\odot}$. \cite{Li2003ApJ...594..987L} found that their best model
required $M_{\mathrm{PAH}}\simeq$~2.2 $\times$ 10$^{-11}$ M$_\odot$
for a PAH size distribution with a minimum radius of
$a_{\mathrm{PAH}}$~=~4.6\,\AA. The discrepancy in the PAH mass may
stem from the difference in the assumed PAH size (we used one single
size whereas \cite{Li2003ApJ...594..987L} used a size distribution for
the PAHs), although we also tested the fit to the PAH features with a
smaller PAH ($a_{\mathrm{PAH}}$~=~3.55\,\AA) and an equally good fit
was obtained with the same mass of PAHs. In our modeling the PAH
opacities were taken into account, especially in the UV wavelengths,
in the continuum radiative transfer, whereas
\cite{Li2003ApJ...594..987L} assumed that the disk was optically thin
to the stellar radiation at all wavelengths.

The disk emission is optically thin vertically and radially from the
optical to the millimeter wavelengths, thus the SED does not depend on
the disk inclination. In addition, the quality of the fit to the
observed photometric points does not vary strongly with the disk
scale-height, as anticipated. The best fit to the SED for each gas
scale-height provides the disk structure for the gas modeling with
{\sc ProDiMo}.

\section{Gas modeling}\label{gas_modeling}

\subsection{Mass accretion rate}\label{mass_accretion_rate}

The mass accretion rate onto \source\ is debated.
\citet{GarciaLopez2006A&A...459..837G} derived from the Br$\gamma$
emission a current mass-accretion rate of 4$\times$10$^{-9}$ M$_\odot$
yr$^{-1}$ using an empirical relation. The derivation of the
mass-accretion rate from the Br$\gamma$ line is not well
calibrated. It is also not known whether \source\ has a
magnetosphere. The double-peaked structure observed in this fast
rotator is different from to what would be expected from a line formed
in a magnetosphere. On the other hand,
\citet{Merin2004A&A...419..301M} did not detect mass-accretion
activity and assumed a lower limit on the accretion rate of 10$^{-11}$
M$_\odot$ yr$^{-1}$. Finally, \citet{Mendigutia2011A&A...535A..99M}
found a mass-accretion rate of 1.3$\times$10$^{-7}$ M$_\odot$ using
the Balmer excess method. It is not clear how the accretion rate has
varied during the lifetime of the disk. In the rest of the paper, we
assume that the disk is passively heated.
\begin{figure*}[!ht]
\centering
\includegraphics[scale=0.5,angle=0]{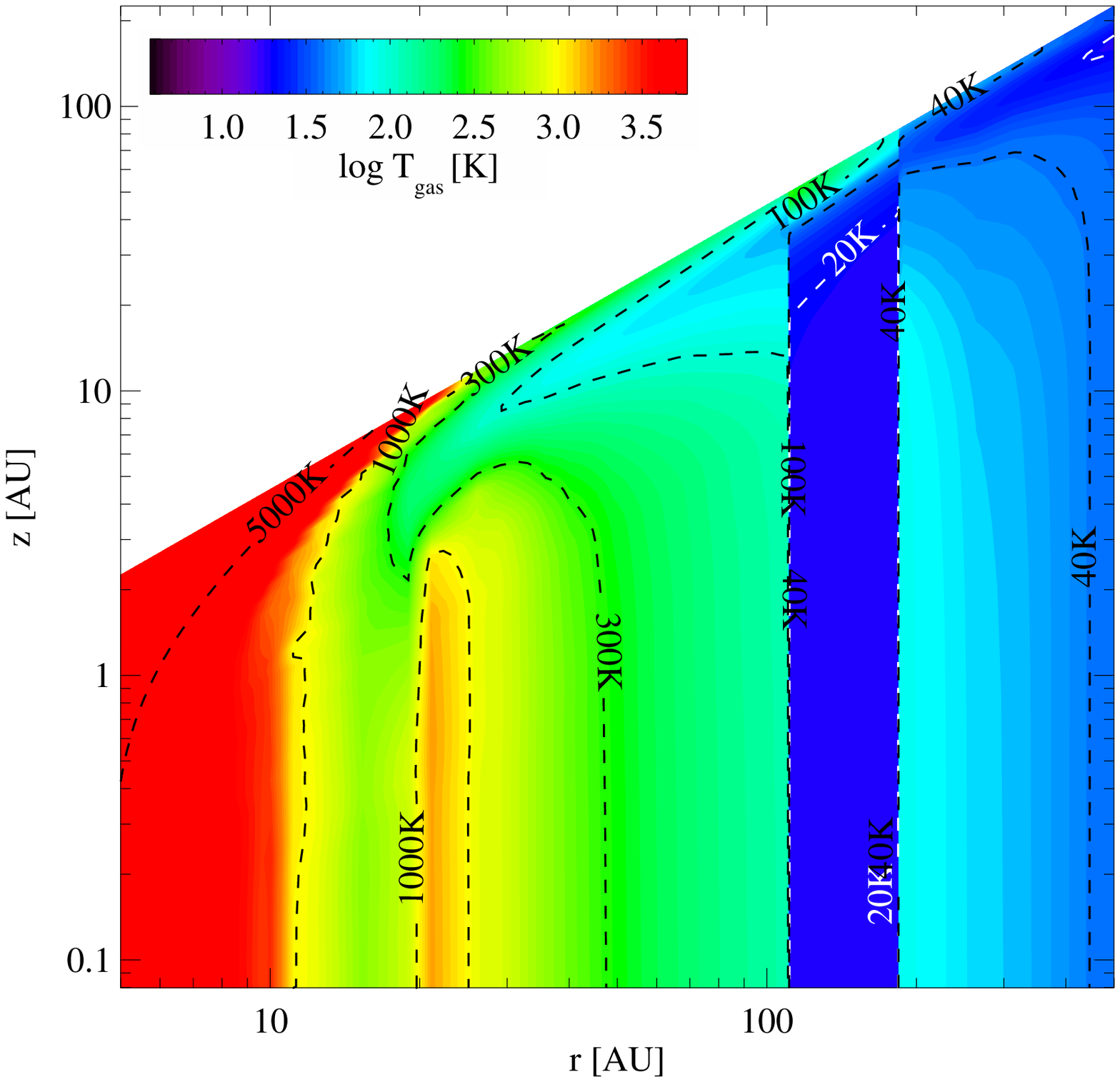}
\includegraphics[scale=0.5,angle=0]{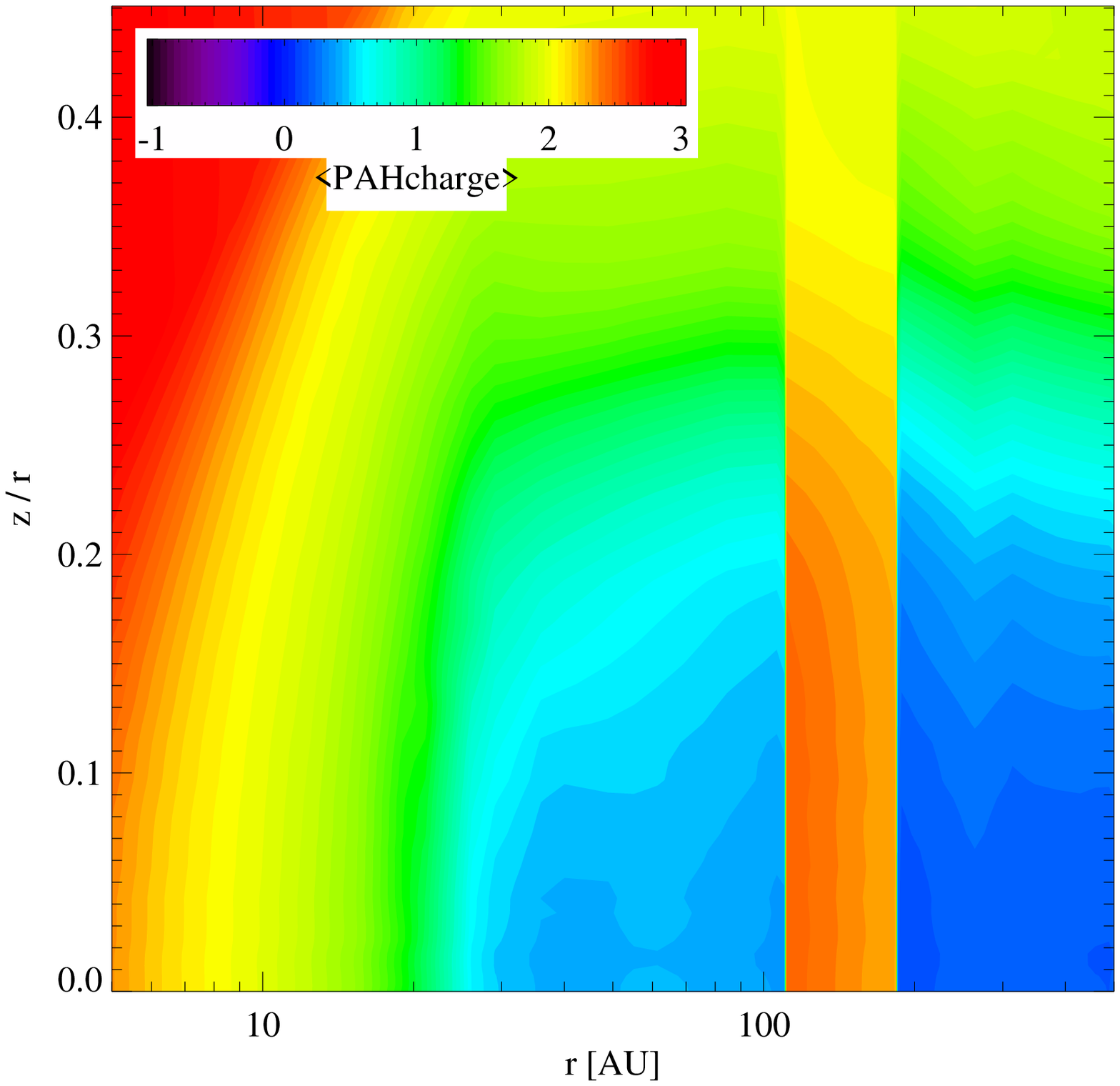}
\includegraphics[scale=0.5,angle=0]{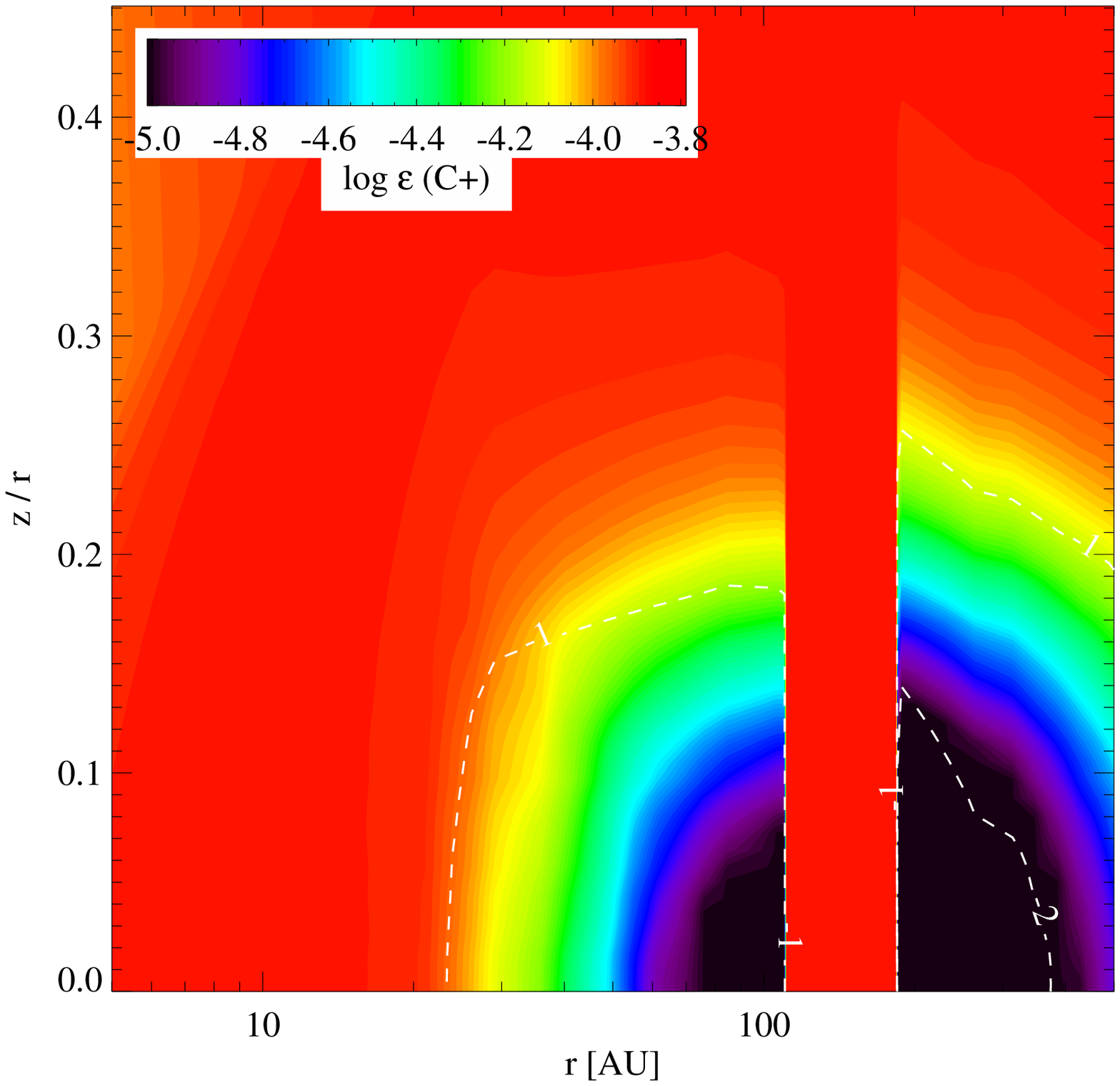}
\includegraphics[scale=0.5,angle=0]{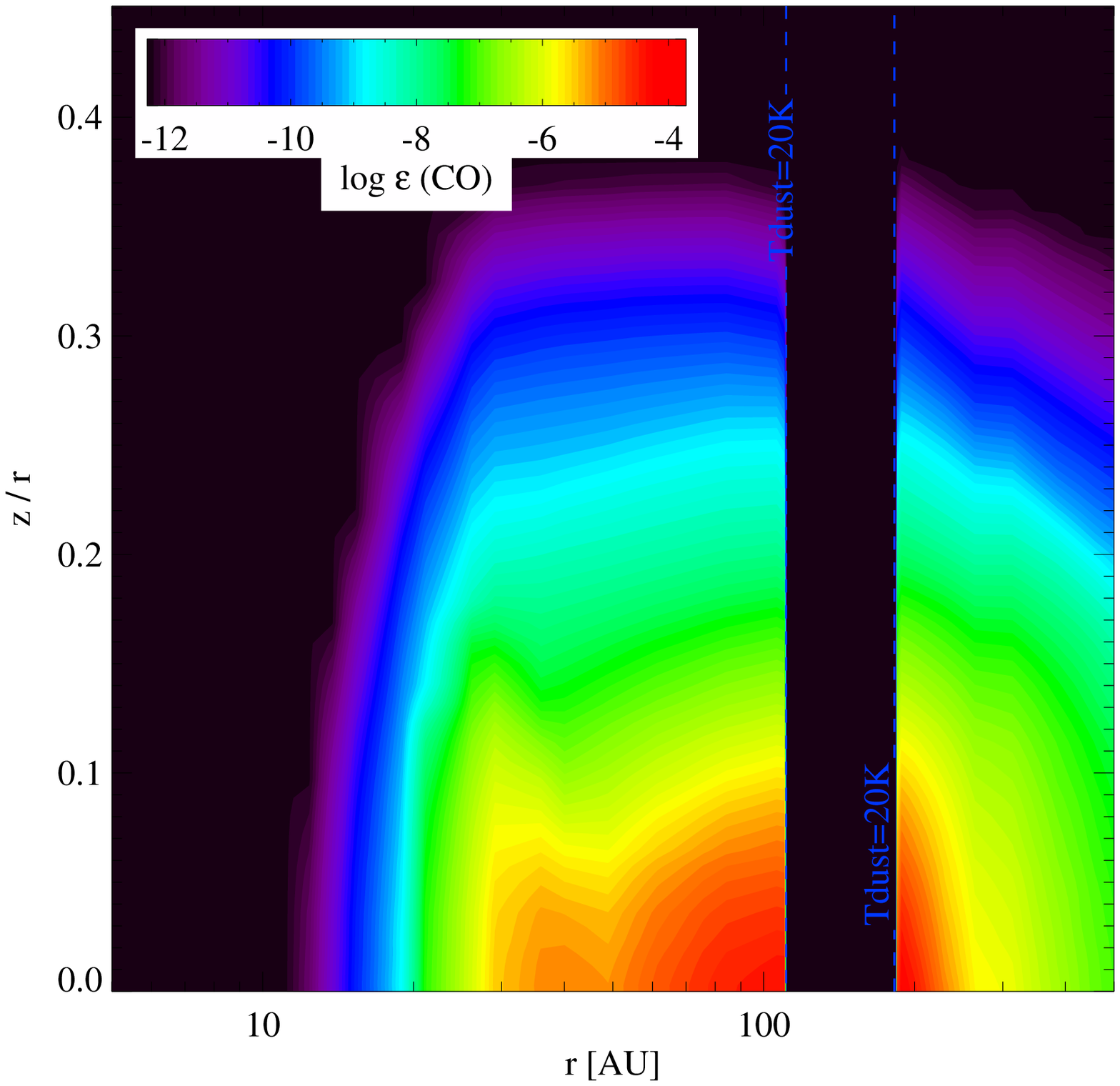}
\caption{The gas temperature structure (upper-left panel), PAH average
  charge (upper-right panel), C$^{+}$ abundance (lower-left panel) and
  the CO abundance (lower-right panel) for the model with $H_0$=5\%
  and a gas-to-dust-mass ratio of 100. The gap between 100 and 185 AU
  is not empty but is filled with a very low density gas. The
    computation of the gas temperature in the low gas density gap is
    not reliable.}
  \label{fig_chemistry}            
\end{figure*}    

\subsection{Chemistry}\label{chemistry}

{\sc ProDiMo} computes the abundance at steady-state of 188 gas and
ice species, including deuterated species, PAHs, and PAH ions (PAH$^-$,
PAH, PAH$^+$, PAH$^{2+}$, PAH$^{3+}$). The photoreaction rates
(photodissociation and photodesorption) were computed from the local UV
field inside the disk and experimental or theoretical cross-sections
\citep{vanDishoeck2008}. The kinetics rates were taken from the
UMIST2006 database \citep{Wooddall2007A&A...466.1197W} with additions
and modifications. PAHs are not formed or destroyed in our chemical
network and participate only in charge-exchange reactions. PAH
photoionization, recombination, and charge-exchange reactions were
added using the rates from \citet{Flower2003MNRAS.343..390F}.

\subsection{Line emission modeling}\label{line_modeling}

For the line observations, we augmented the {\sc Herschel} data with
JCMT $^{13}$CO $J$=3--2 observations published by
\citet{Dent2005MNRAS.359..663D}. After constraining the dust disk mass
from the SED, we ran a series of models with the thermo-chemical
code {\sc ProDiMo} \citep[a detailed description is given
in][]{Woitke2009A&A...501..383W,Kamp2010A&A...510A..18K,thi2011MNRAS.412..711T}.
The effects of X-rays on the gas properties in protoplanetary disks are
discussed in \cite{Aresu2012A&A...547A..69A,Aresu2011A&A...526A.163A}
and \cite{Meijerink2012A&A...547A..68M}. The collisional data for the
line transfer were taken from the {\it Leiden-Lambda} database
\citep{Schoier2005A&A...432..369S}. The original references for the
experimental or theoretical rates are given in
Appendix~\ref{collisional_rates}. 

\begin{figure*}[!ht]
\centering
\includegraphics[scale=0.5,angle=0]{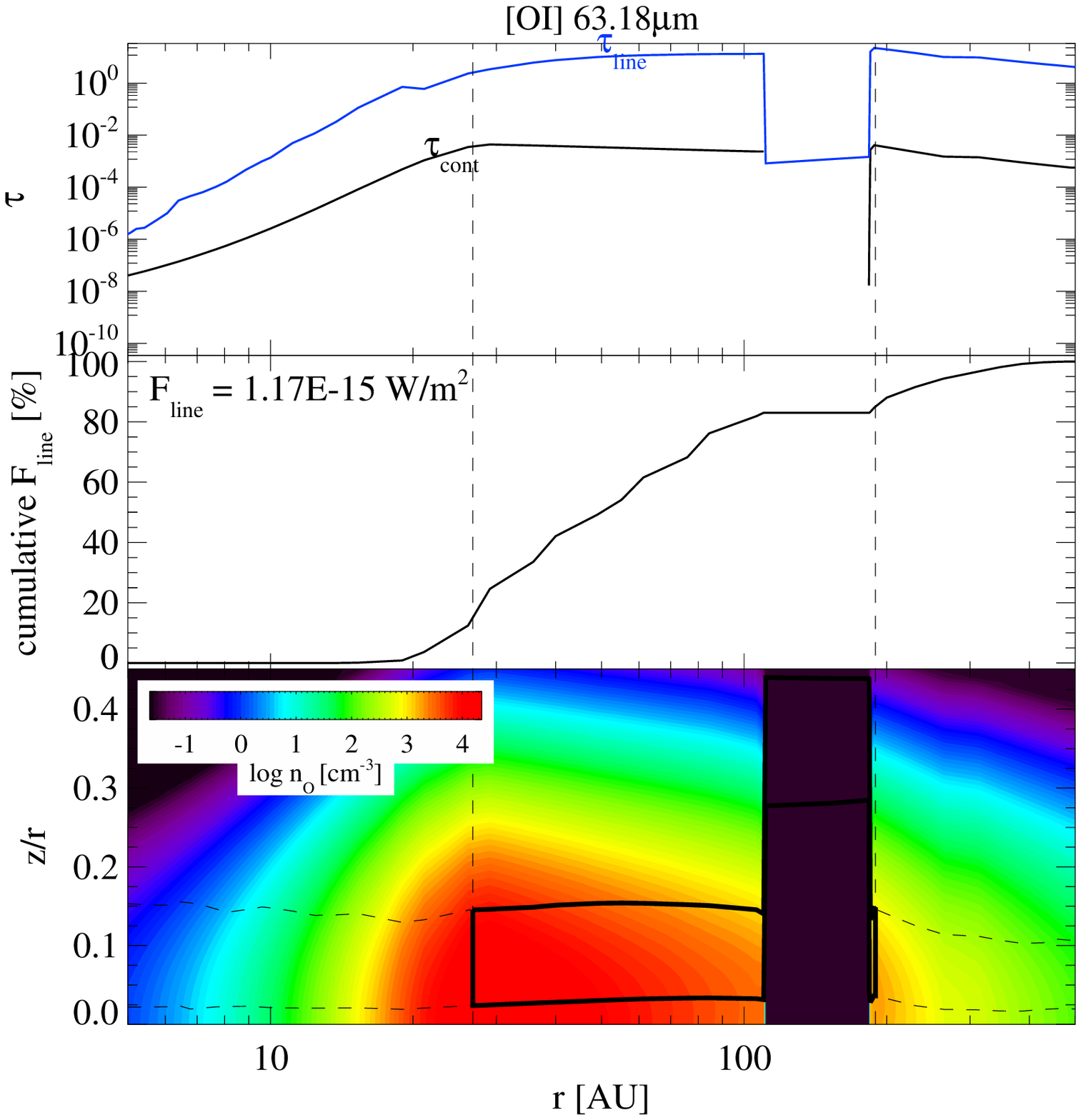}
\includegraphics[scale=0.5,angle=0]{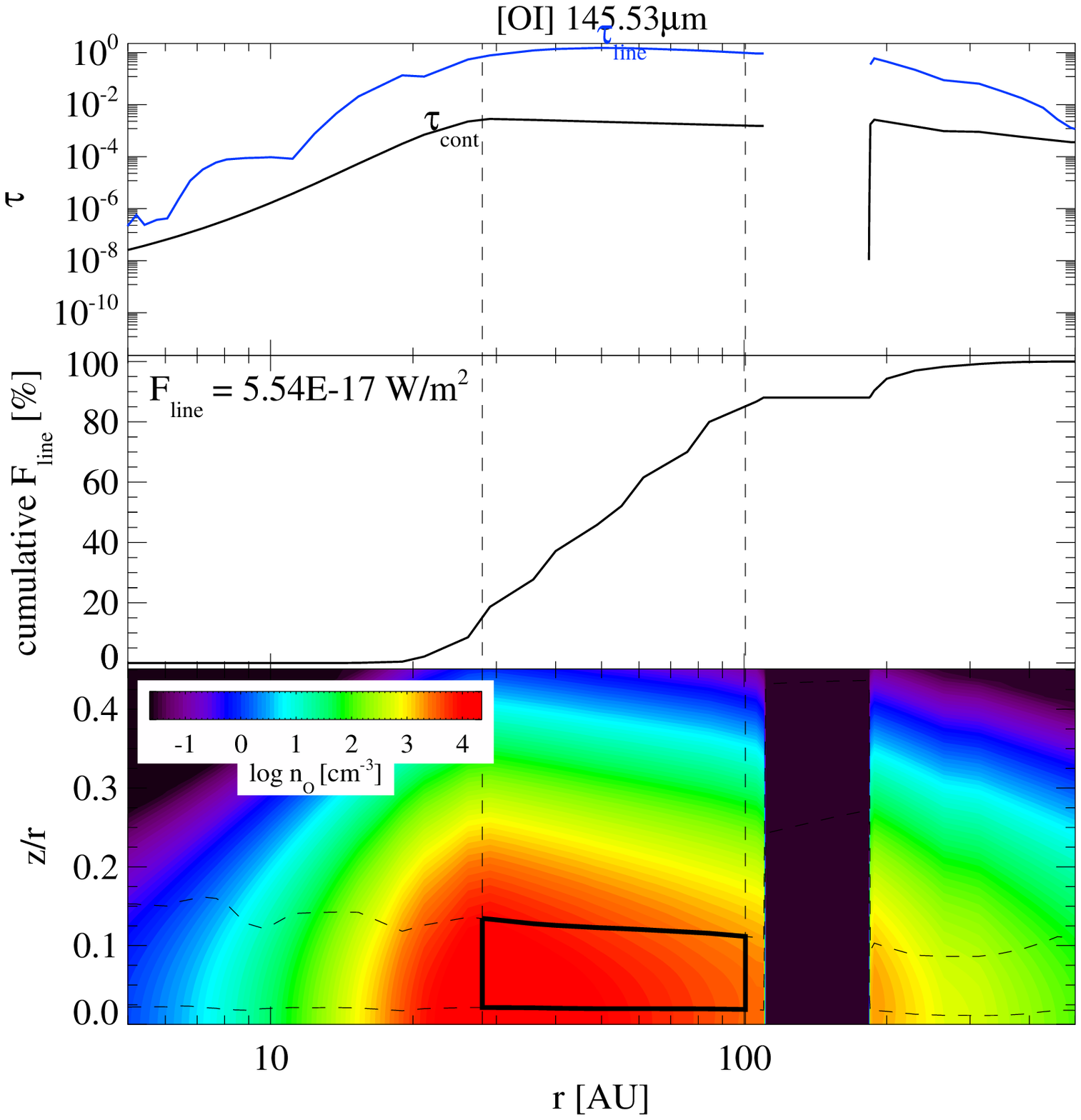}
\includegraphics[scale=0.5,angle=0]{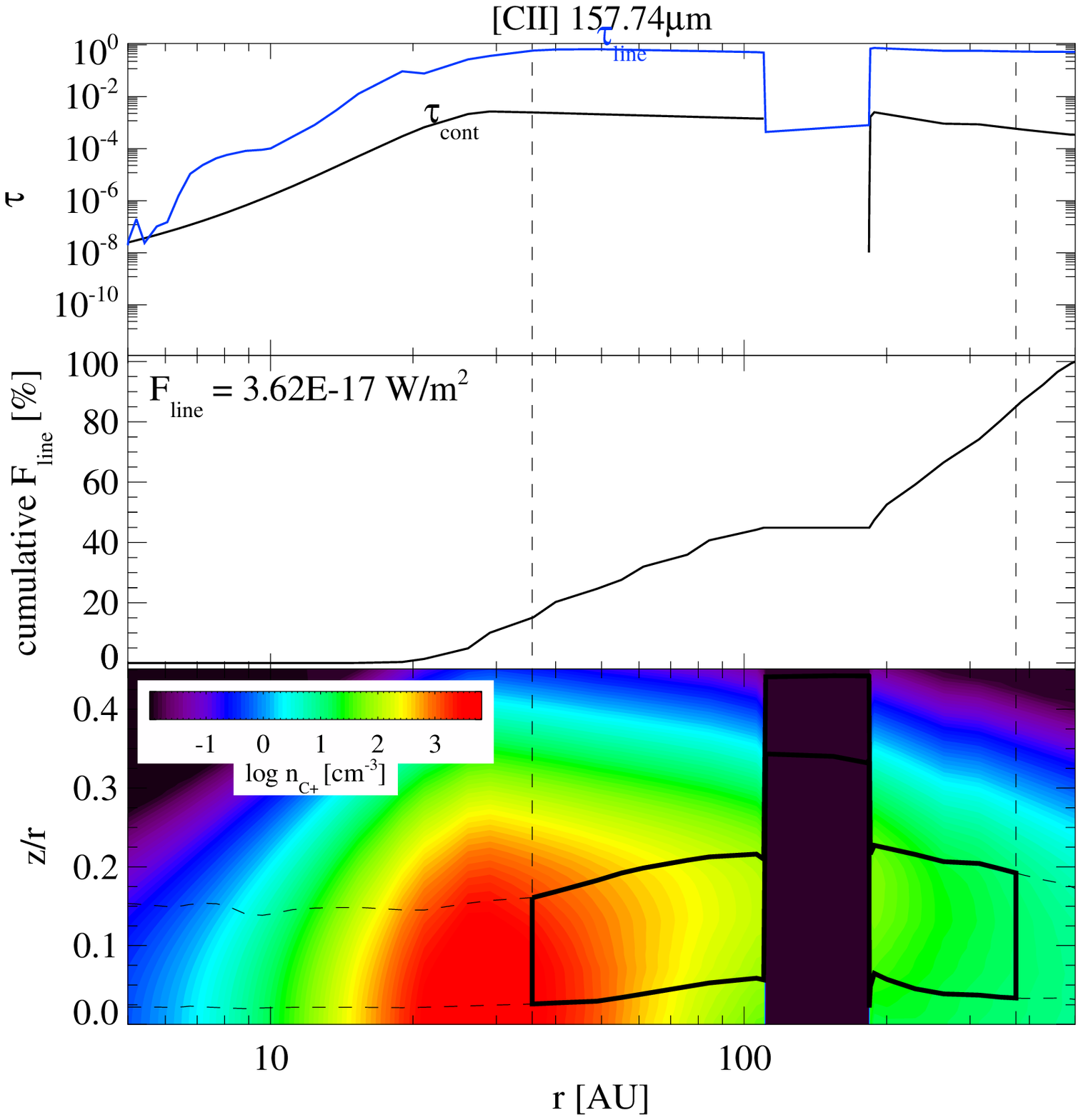}
\includegraphics[scale=0.5,angle=0]{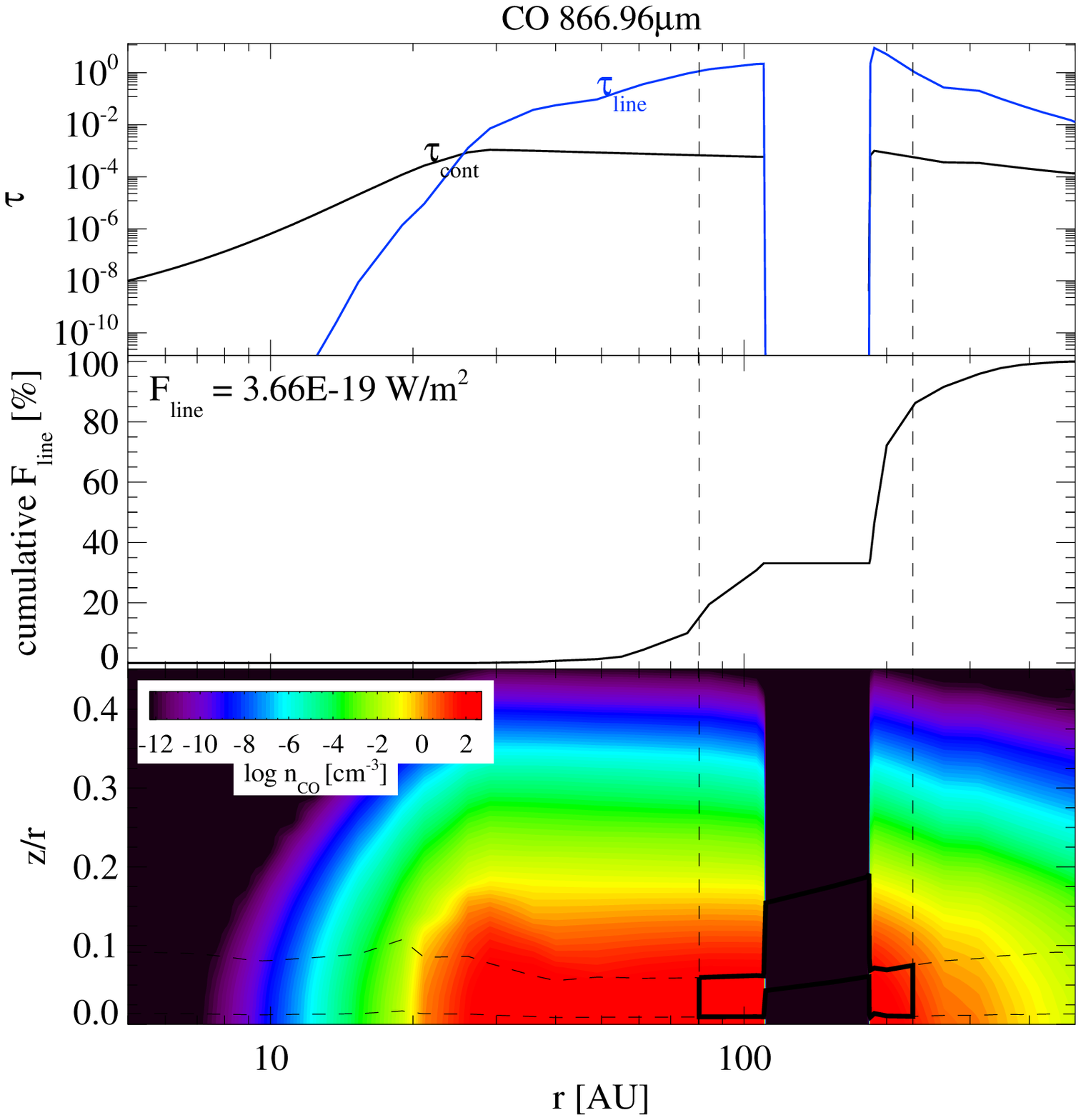}  
\caption{The \OIfs\ emission at 63 microns (upper-left panel), at 145
  microns (upper-right panel), \CII\ emission at 157 microns
  (lower-left panel) and the CO 3-2 emission (lower-right panel) for
  the model with $H_0$=5\% and a gas-to-dust-mass ratio of 100. The
  upper part in each panel shows the line center and continuum
  vertical optical depth as function of the disk radius. The middle
  part is the cumulative flux as function of the radius. The lower
  part shows the species volume density in cm$^{-3}$. The line
    fluxes are computed by ray-tracing from the opposite side of the
    disk toward the observer. The black boxes in the lower panels
    represent the regions where the vertical cumulative fluxes are at
    15\% to 85\% of the fluxes that are emitted for half a disk.  The
  fluxes are given for a disk seen face-on. For a given disk
  inclination, optical depth will change the emitted fluxes.}
  \label{fig_line_emissions}          
\end{figure*}          
\begin{figure*}[!ht]  
\includegraphics[scale=0.38,angle=90]{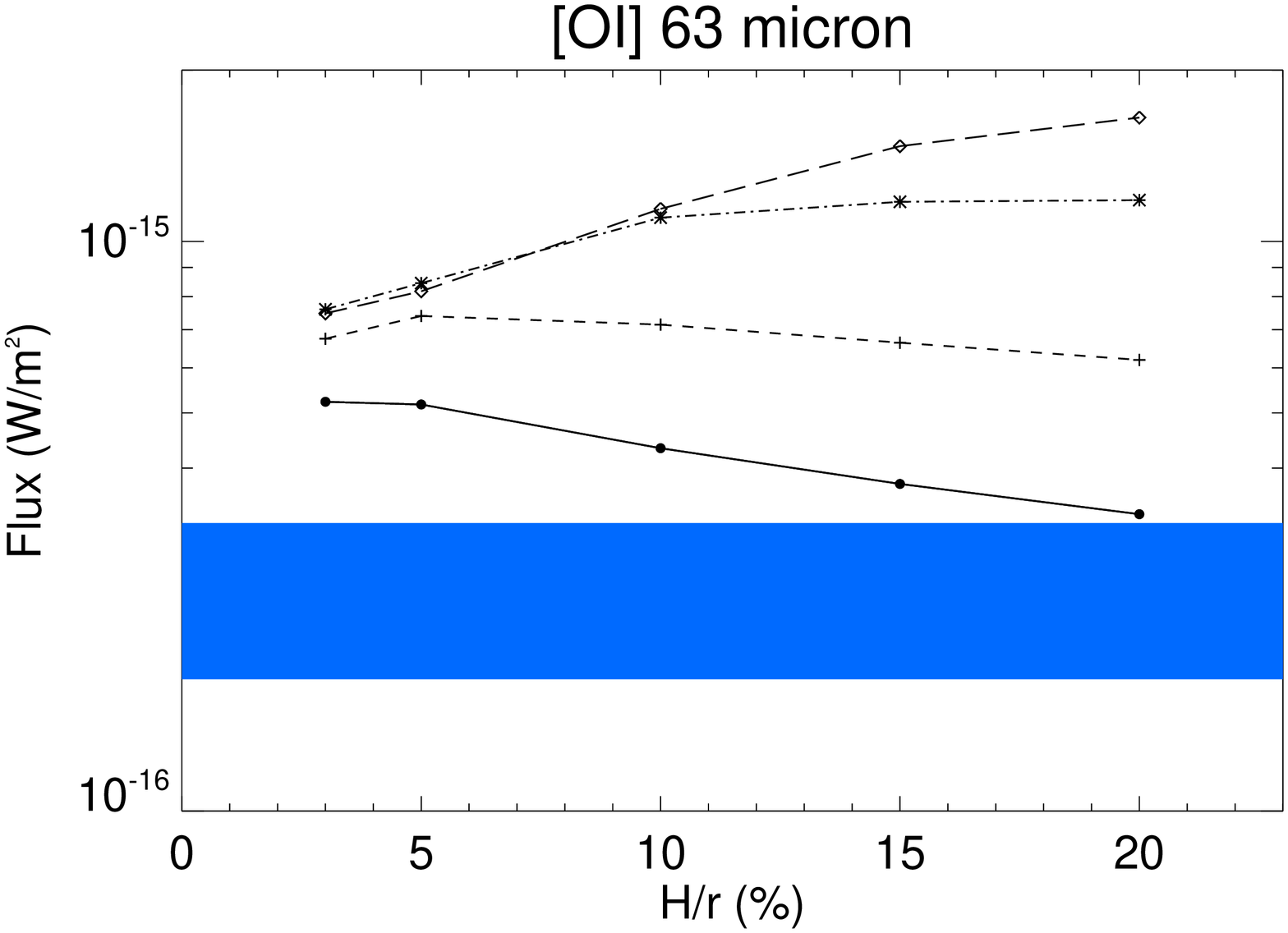}  
\includegraphics[scale=0.38,angle=90]{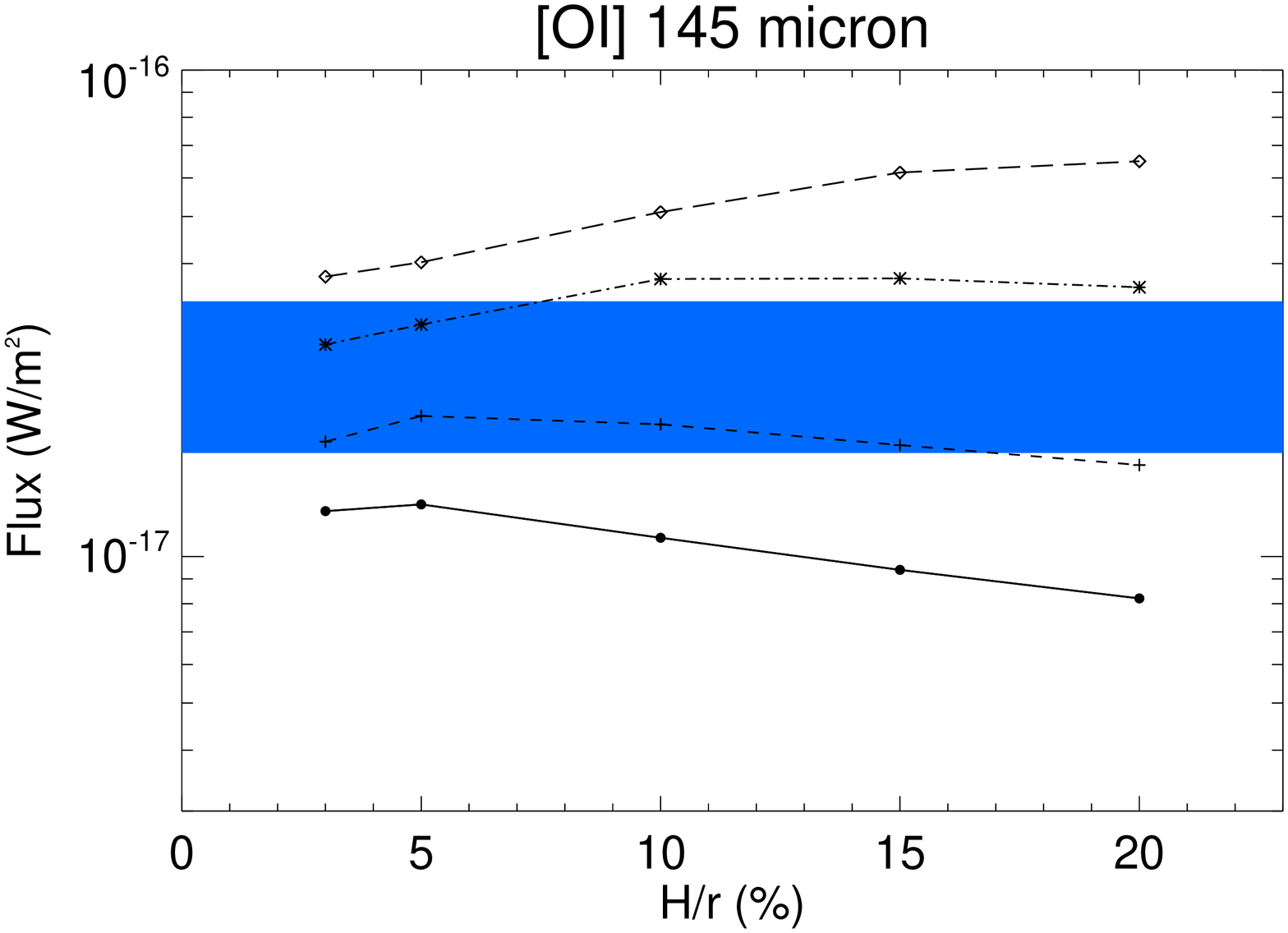}
\includegraphics[scale=0.38,angle=90]{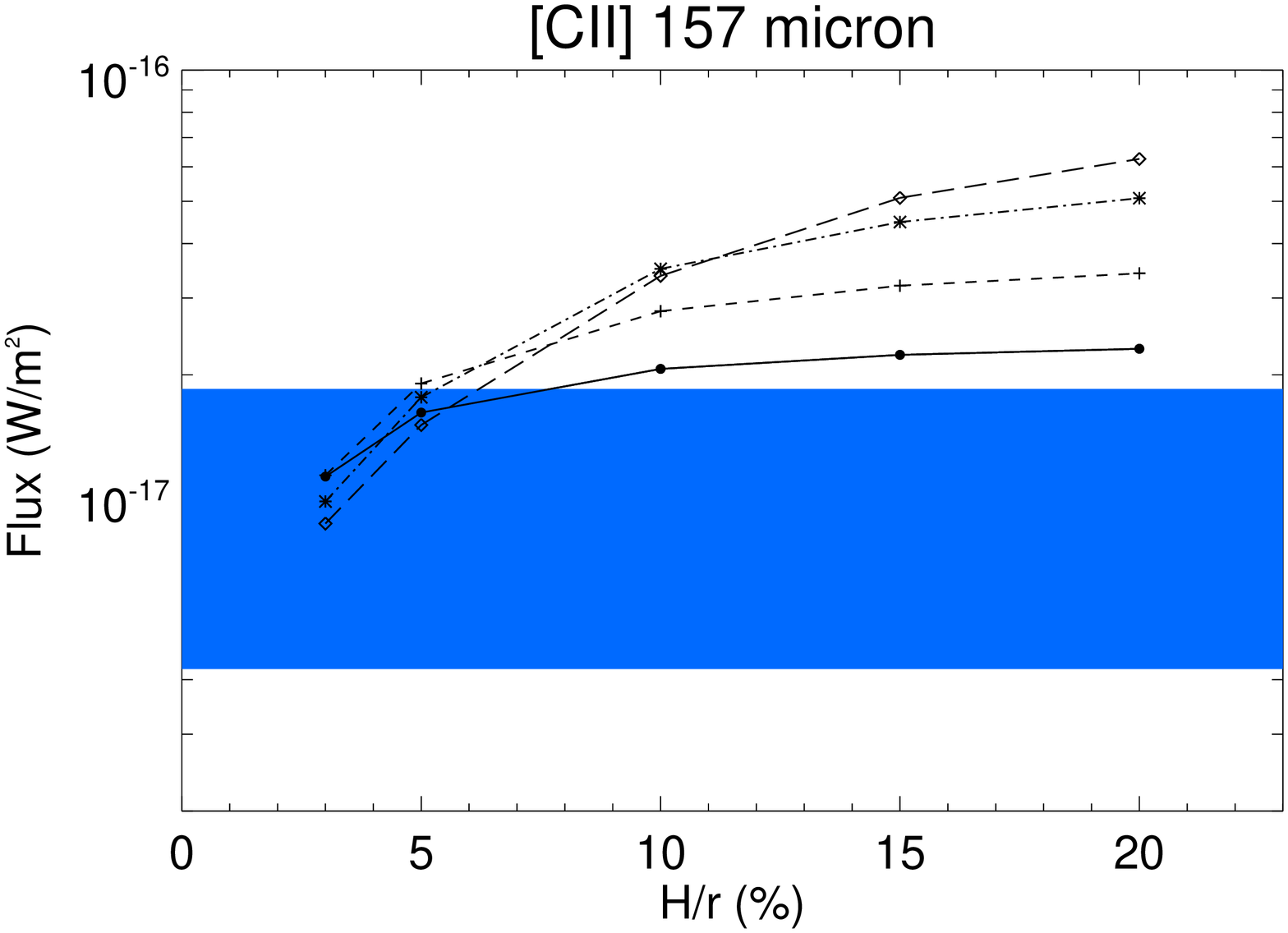}
\includegraphics[scale=0.38,angle=90]{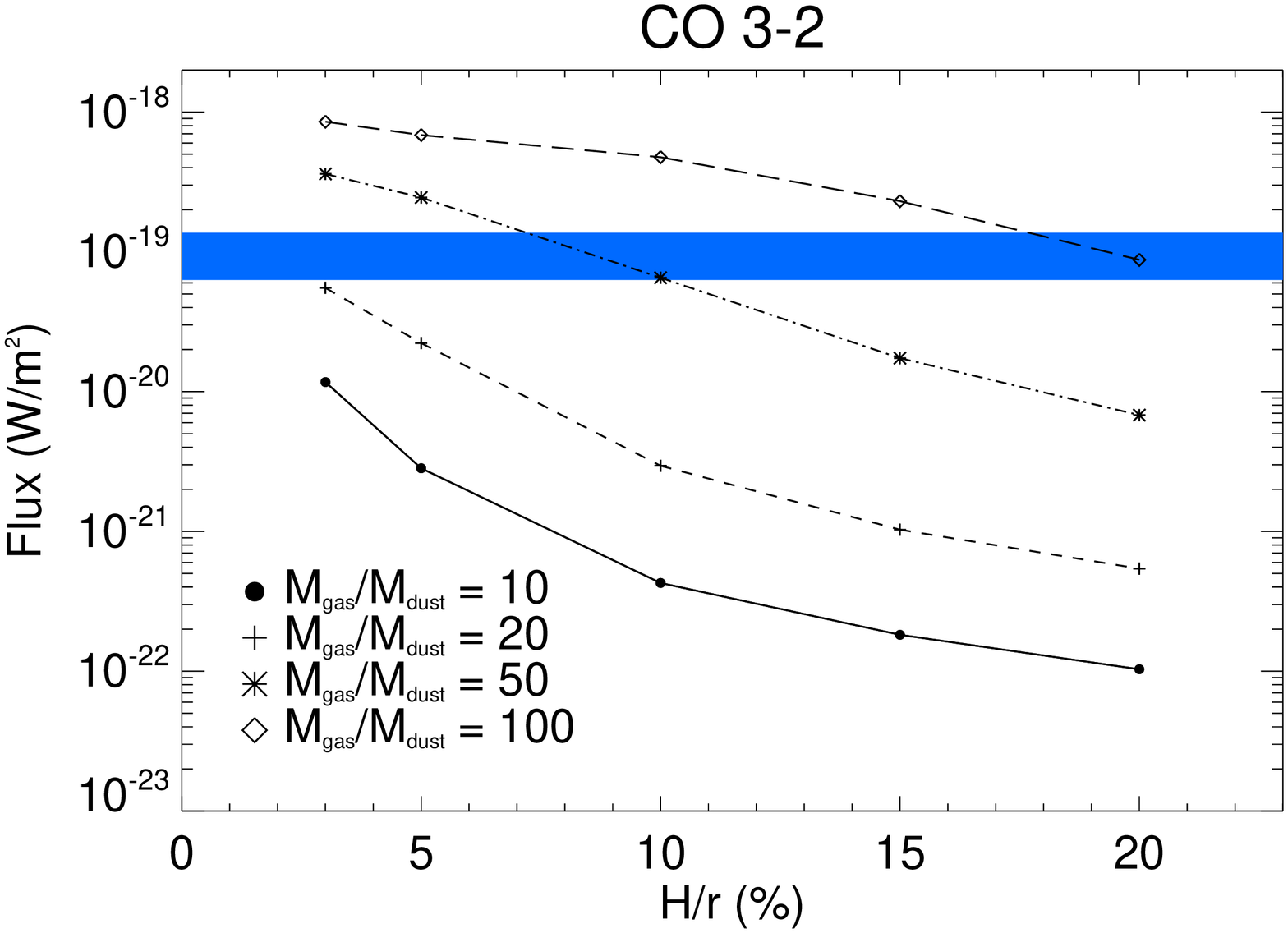}
\caption{Predicted [\OI]\ 63 and 145 micron, [\CII]\ 157 micron
  fine-structure and CO 3-2 fluxes as a function of the gas
  scale-height parameter for four different gas-to-dust-mass ratio.
  The observed fluxes are overplotted in blue boxes with the
  3~$\sigma$ error and 30\% calibration error taken into account.
    No model fits the [\OI]\ 63 micron line.}
  \label{fig_full_models}                  
\end{figure*}    

\subsection{Gas kinetic temperature}

The gas kinetic temperature was computed by balancing heating and
cooling processes. Line profiles were computed by non-LTE radiative
transfer within {\sc ProDiMo}.

The disk was assumed to be passively heated with no viscous heating
($\alpha$=0). Heating processes include photoelectric heating (PAHs
and dust grains), fluorescence pumping followed by collisional
de-excitation, gas-grain accommodation (which can also be a cooling
agent), H$_2$ formation on grain surfaces, cosmic-ray heating, and
X-ray (with a luminosity of $L_{\mathrm{X}}$=10$^{28}$ erg
s$^{-1}$). For \source, we assumed that the X-ray-heating is
negligible compared with the heating via PAH photoelectric effects
used in \citet{Meijerink2012A&A...547A..68M} and
\citet{Aresu2012A&A...547A..69A,Aresu2011A&A...526A.163A}.

The profile of the $^{12}$CO $J$=3--2 emission line constrains the
disk turbulent velocity $v_{\mathrm{turb}}$ to below 0.2 km
s$^{-1}$. We adopted a typical value of 0.15 km s$^{-1}$.

The outer disk is irradiated by direct and scattered stellar photons
as well as by interstellar UV photons ($G_0$=$\chi$=1). The free
parameters of the gas simulations are the disk gas mass
$M_{\mathrm{gas}}$ (or equivalently the gas-to-dust-mass ratio since
the total dust mass is fixed in the gas modeling), the fraction of
PAHs in the disk with respect to the interstellar abundance
$f_{\mathrm{PAH}}$, and the cosmic-ray flux $\zeta$ (=5 $\times$
10$^{-17}$ in the interstellar medium). Observations show that PAHs
are depleted by at least a factor of 10 ($f_{\mathrm{PAH}}=0.1$) in
disks with respect to the interstellar abundance in protoplanetary
disks \citep{Geers2006A&A...459..545G}.  We adopted the PAH abundances
derived from the total PAH masses (Table~\ref{PAH_abundance}). The gas
is mostly heated by photoelectrons ejected from PAHs, whose abundance
is constrained by fitting the PAH IR features. In low-mass disks
irradiated by a luminous central objects, fluorescence pumping by
direct or scattered stellar light can be significant. Electronic
levels of OI, CII, and CO were taken into account. For CO, we included
50 rotational levels for each of the nine vibrational levels and two
electronic levels (the ground- and $A$ electronic level). The
collisional rates were computed according to
\citet{thi2013A&A...551A..49T}.

\subsection{Chemistry and line modeling results}~\label{line_results}

We ran a series of models for each value of the gas scale-height at
100~AU $H_0$: 3, 5, 10, 15, and 20~AU. In each series, the gas mass
was allowed to vary such that the gas-to-dust-mass ratio was 10, 20,
50, and 100, the solid mass was kept constant. We chose to keep a
constant gas-to-dust-mass ratio throughout the disk, in particular, we
did not account for dust settling.

We modeled the emission of the three fine-structure lines and the CO
$J$=3--2 line to constrain the disk gas mass. An example of the gas
temperature, PAH charge, the C$^+$ and CO abundance for the disk with
a 5\% scale-height and a gas-to-dust ratio of 100 is shown in
Fig.~\ref{fig_chemistry}. The gas temperature shows a
  high-temperature feature in the 20 and 30~AU region, which
  corresponds to the location of the maximum surface density.

  The level populations were computed at NLTE and line profiles are
  generated using ray-tracing. The line fluxes in W m$^{-2}$ were
  obtained by integrating the line flux profiles. The locations of the
  fine-structure and CO $J$=3--2 emission are shown in
  Fig.~\ref{fig_line_emissions}. A summary of all model results
  is plotted in Fig.~\ref{fig_full_models}.

\begin{figure}[!ht]  
\resizebox{\hsize}{!}{\includegraphics[scale=1,angle=90]{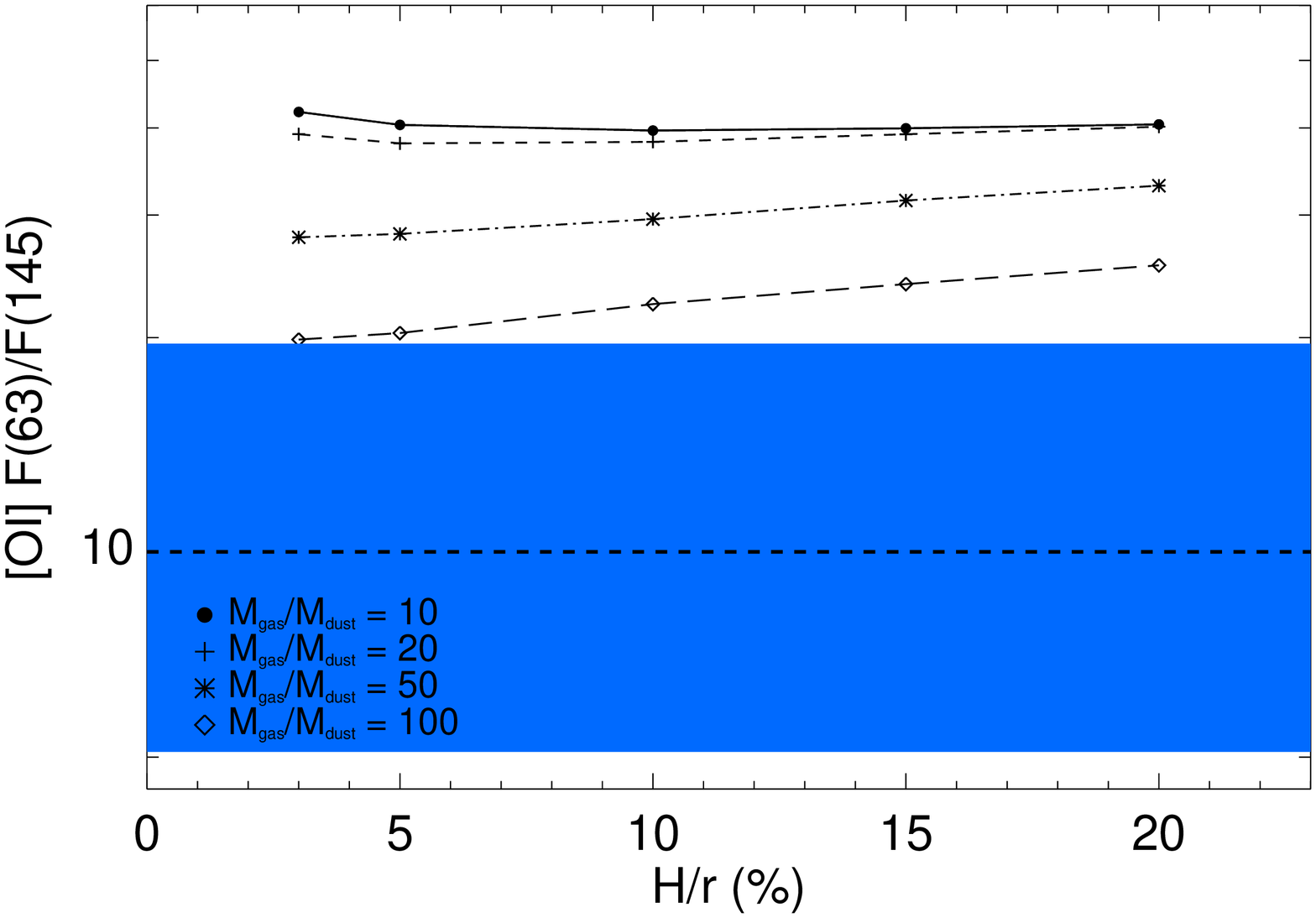}}  
\caption{\OIfs\ 63$~\mu$m/145$~\mu$m line flux ratios as function of
  the disk gas scale-height for different gas-to-dust-mass ratios. The
  observed ratio is shown by a horizontal dash line and the
  uncertainties on the ratio by a blue box.}\label{OI145_63ratios}
\end{figure}  
The gas is mostly composed of molecular hydrogen. Despite the low
optical depth in both the radial and vertical direction,
photodissociation of H$_2$ is overcome by its rapid formation on grain
surfaces and self-shielding at steady-state.

Atomic oxygen is the most abundant oxygen-bearing species throughout
the disk. Carbon is ionized at the disk surfaces and in the form of CO
closer to the midplane (Fig.\ref{fig_chemistry}). In the most massive
disk models (M$_{\mathrm{gas}}>$10$^{-3}$ M$_\odot$), CO becomes the
dominant carbon-bearing species in the midplane (3$\times$10$^{-4}$
the H$_2$ abundance). 
\begin{figure}[!ht]  
\resizebox{\hsize}{!}{\includegraphics[scale=1,angle=90]{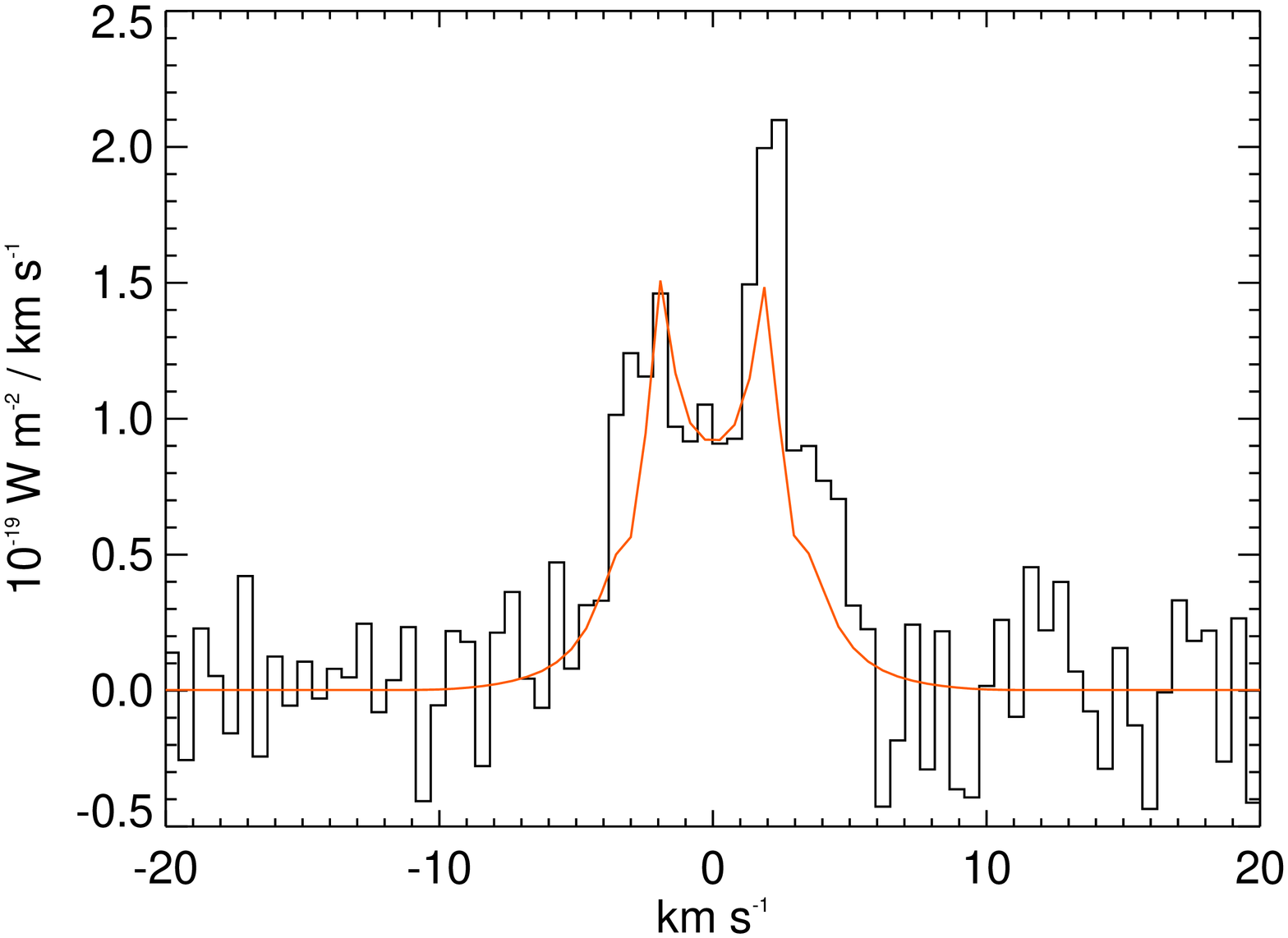}}  
\caption{Observed CO $J$=3--2 line profile (histogram black line)
   compared with the disk model-predicted profile (continuous red
  line). The model has an opening angle of 5\% and a gas-to-dust-mass
  ratio of 100).}\label{CO_3_2_profile}
\end{figure}  
\begin{figure}[!ht]  
\resizebox{\hsize}{!}{\includegraphics[scale=1,angle=90]{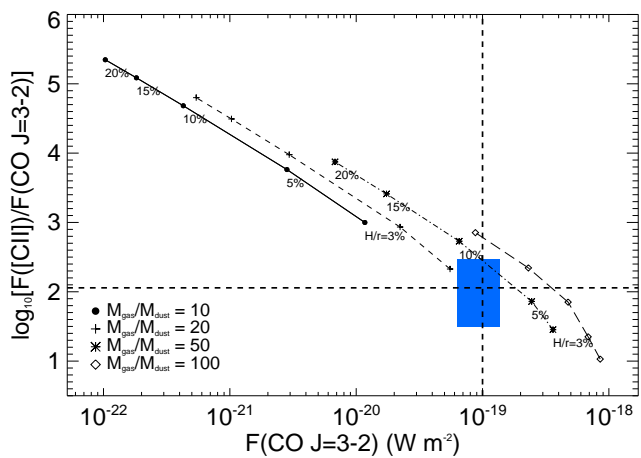}}  
\caption{Observed \CIIfs\ / CO $J$=3--2 line flux ratios as a
  function of the CO $J$=3--2 line flux for different models with
  varying opening angles $H/r$ (3\%, 5\%, 10\%, 15\%, and 20\% as
  indicated on the plot) and disk gas-to-dust-mass ratios. The blue box
  encircles the observed ratio and flux, including the uncertainties.
}\label{CII_CO_3_2_CO_3_2ratios}
\end{figure}  
\begin{figure}[!ht]  
\resizebox{\hsize}{!}{\includegraphics[scale=1,angle=90]{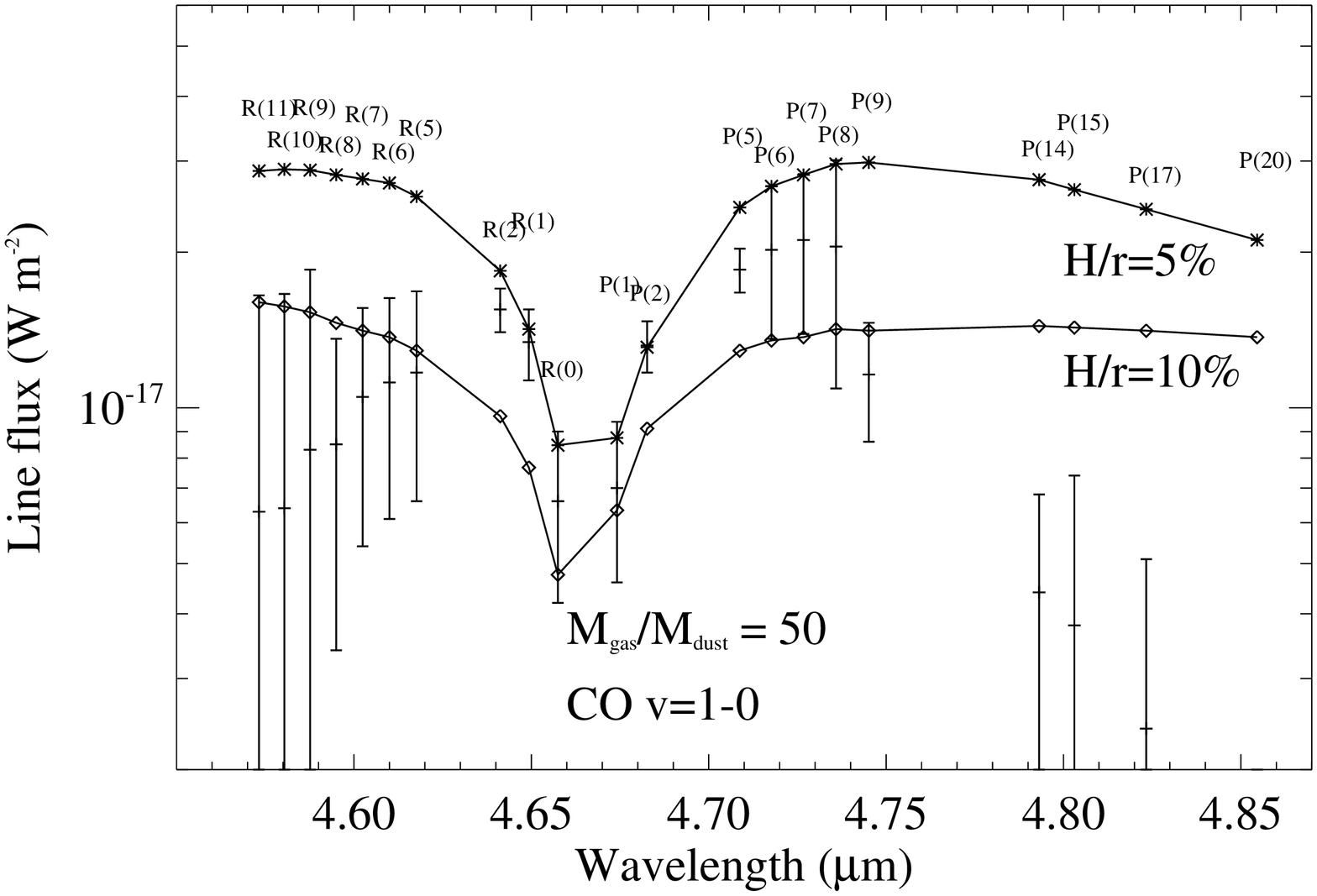}}  
\caption{Some modeled CO fundamental ro-vibrational fluxes
  compared with the observed fluxes taken from
  \citet{Brittain2007ApJ...659..685B}. The error bars are 3~$\sigma$
  uncertainties. The identification of the lines are labelled. The two
  models have a gas-to-dust-mass ratio of 50. The opening angles are 5\%
  and 10\%.}\label{CO_rovib}
\end{figure}  
The gas is heated by PAH photoelectrons despite the low PAH abundances
compared with the interstellar abundance of
$\sim$~3$\times$~10$^{-7}$. The PAHs are slightly positive because the
electron recombination is almost as fast as the photoejection of
electrons (Fig.~\ref{fig_chemistry}). The average charge of the PAHs
depends on the gas density. At low densities the photoejection rate
dominates the recombination rate. CO ro-vibrational absorptions of the
stellar and dust-emitted IR photons also contribute to the heating of
the gas in the inner disk.

The gas is cooled by \OIfs\ 63 $\mu$m and CO ro-vibrational line
emissions, two of the main gas components of the disk in addition to
H$_2$. The gas and the dust are not coupled thermally, which is
different from more massive disks. The gas temperature in the inner
disk is between 100 and 300~K, and in the outer disk it is between 40
and 100~K.

The two \OIfs\ lines are optically thin or weakly thick for the most
massive disks (Fig.~\ref{fig_line_emissions}). The \OIfs\ lines probe
the inner and outer disk. In the upper left panel of
Fig.~\ref{fig_full_models}, all the \OIfs\ 63$~\mu$m fluxes are higher
than the observed value. The 145~$\mu$m flux varies with the
gas-to-dust-mass ratio (i.e., the disk gas mass), but is not strongly
dependent on the gas scale-height. The 145~$\mu$m flux can be
reproduced by disk models for all gas scale-height $H_0$ and
gas-to-dust-mass ratio of 20 to 50. The \OIfs\ 63$~\mu$m/145$~\mu$m
line flux ratios are shown in Fig.~\ref{OI145_63ratios}. The model
ratios are higher than 10 and decrease with increasing total disk gas
mass, which reflects the effect of the line optical depth
\citep{Liseau2006A&A...446..561L}.

The \CIIfs\ line is optically thin and the flux increases with
radius. \CIIfs\ emission is emitted both by the inner and outer disk
($\sim$~70\% by the inner disk and $\sim$~30\% by the outer disk, see
Fig.~\ref{fig_line_emissions}). As the disk becomes more massive, more
carbon is converted into CO and the disk becomes cooler. Ionized
carbon is excited in gas at $\sim$~100~K. The \CIIfs\ flux first
starts to increase with higher disk gas mass, but then plummets for
disk gas masses greater than a few 10$^{-3}$ M$_{\odot}$
(Fig.~\ref{fig_full_models}).

The CO $J$=3--2 flux increases with increasing disk gas mass. CO
$J$=3--2 emission is generated mostly by the outer disk (70\% for
$R>$~185~AU, Fig.~\ref{fig_line_emissions}) with significant
contribution from the inner disk. The CO $J=$3--2 flux varies
dramatically with the gas scale-height for low-mass disks. Reaction
rates forming CO vary as $n_{\mathrm{H}}^2$, where $n_{\mathrm{H}}$ is
the gas density. In disks with large scale-heights, the gas is more
diluted and the CO formation rate and self-shielding cannot compensate
for the photodissociation.  As a result, the \CIIfs\ flux increases
with increasing gas scale-height. An example of a predicted line
profile compared with the observed spectrum obtained by
\cite{Dent2005MNRAS.359..663D} is shown in
Fig.~\ref{CO_3_2_profile}. The \CIIfs\ / CO $J$=3--2 line flux ratio
as a function of the CO $J$=3--2 line flux as plotted in
Fig.~\ref{CII_CO_3_2_CO_3_2ratios} is very sensitive to the
gas-to-dust-mass ratio and gas scale-height. This plot suggests a
gas-to-dust-mass ratio of $\sim$50 and a scale-height between 5 and
10\%.
   
All together, the fine-structure and CO $J$=3--2 lines constrain the
disk gas-to-dust-mass ratio between 50 to 100 and a gas scale-height
of 5-10\%. Especially the low value of the observed \OIfs\
63$~\mu$m/145$~\mu$m line flux ratio suggests a disk massive enough
such that the \OIfs\ lines become marginally optically thick.

In low-mass disks, the accuracy of CO photodissociation self-shielding
factors determines the CO abundance. In all our disk models, CO is
entirely in the gas-phase. The dust temperature in the outer disk is
$\sim$40--60~K, which is higher than the sublimation temperature of CO
ice. Therefore the major carbon-bearing species are ionized atomic
carbon, neutral atomic carbon, and CO.

Finally, we compared the CO fundamental ro-vibrational ($v$=1-0) line
flux predictions with the observed values of
\citet{Brittain2007ApJ...659..685B} in Fig.~\ref{CO_rovib}. We did not
use the CO ro-vibrational lines to constrain the disk parameters. In
general, the modeled and observed line fluxes differ within a factor
two from each other for disk models with a gas-to-dust-mass ratio
between 50 and 100 and a gas scale-height between 5\% and 10\%. The CO
ro-vibrational lines are emitted between 10~AU and 50~AU consistent
with the constraints given by the spatially resolved CO ro-vibrational
observations of \citet{Goto2006ApJ...652..758G}. The CO ro-vibrational
emission surface brightness is at its maximum at the location of the
maximum temperature region between 20~AU and 30~AU.

In summary, the fine-structure and CO $J$=3--2 line fluxes suggest
that \source\ currently has a gas mass of (2.5-5)~$\times$~10$^{-4}$
M$_\odot$.  The detailed modeling suggested gas masses that are
consistent with the values found by the simple interpretation in
Sec.~\ref{simple_analysis}. However, the gas temperatures in the
detailed models are higher than those derived from the simple
analysis.

\subsection{Low oxygen abundance in the disk around  \source ?}

One of the fixed parameters that was not varied is the oxygen
abundance. Large amounts of oxygen can be locked as water ice onto
km-sized planetesimals and decrease its gas phase abundance by a
significant amount \citep{Ciesla2006Icar..181..178C}. A low oxygen
abundance has been discussed as possible in the inner region of
T~Tauri disks \citep{Najita2013ApJ...766..134N}. We tested the
influence of the oxygen abundance by decreasing its value by a factor
two, four, and ten in two of the models. The results are summarized in
Table~\ref{Low_oxygen_abundance}. As expected, the line fluxes of
oxygen-bearing species decrease while the \CII\ line flux
increases. However, the \OIfs\ 63$~\mu$m/145$~\mu$m line flux ratio
increases with decreasing elemental oxygen abundance as the 63 micron
line becomes optically thinner. It is not clear whether a low oxygen
elemental abundance can reduce the \OIfs\ 63 micron flux without
affecting the other fluxes.


\begin{table*}
\begin{center}
\begin{tabular}{llllllllllll}
                  \toprule
Line & & \multicolumn{4}{c}{Model 1} & &\multicolumn{5}{c}{Model 2}\\
\cline{2-6}
\cline{8-12}\\
or line ratio     & \multicolumn{1}{c}{Observations} & \multicolumn{1}{c}{Std.} & \multicolumn{1}{c}{[O]/2} & \multicolumn{1}{c}{[O]/4} & \multicolumn{1}{c}{[O]/10} & &\multicolumn{1}{c}{Std.} &\multicolumn{1}{c}{self-shield.} & \multicolumn{1}{c}{[O]/2} & \multicolumn{1}{c}{[O]/4} & \multicolumn{1}{c}{[O]/10}\\
     &  \multicolumn{11}{c}{(10$^{-18}$ W m$^{-2}$)}\\   
\noalign{\smallskip}   
\hline
\OIfs\  $^3$P$_1 \rightarrow ^3$P$_2$        &   245.3~$\pm$~4.8  & 1217  & 1100 & 937  & 640 & &748   &569  & 653 &556  & 421\\
\OIfs $^3$P$_0 \rightarrow ^3$P$_1$          &   24.9~$\pm$~1.4   & 67.2  & 40.9 & 24.6 & 12.3 & &37.6 & 28.6 & 19.4 &11.9 & 7.2\\
\CIIfs\  $^2$P$_{3/2} \rightarrow ^2$P$_{1/2}$ &   11.4~$\pm$~1.8   & 39.3  & 41.9 & 46.2 & 52.1 & &9.1  & 6.7 & 10.5 &12.1 & 13.5\\
$^{12}$CO $J$=3--2                           &   0.1~$\pm$~0.008  & 0.79  & 0.49 & 0.27 & 0.08 & &0.85 & 1.46 & 0.68 & 0.47 & 0.18\\
\noalign{\smallskip}   
\hline
\noalign{\smallskip}   
\OIfs\ 63/145                               &  9.8              &  18.1 &  26.9 & 38.1 & 52.0 & & 19.9 & 19.9 &33.6 &46.7 & 58.5 \\
\CIIfs\ / $^{12}$CO $J$=3--2                 &  114              &  49.7 & 85.5 & 171.1 & 651.2 & & 10.7 & 4.6 & 15.4 & 25.7 & 75 \\
\noalign{\smallskip}     
\bottomrule
\end{tabular} 
\caption{Line fluxes from disk models: Model 1 with an opening angle
  of 10\% and Model 2 with an opening angle of 3\%. Both models have a
  gas-to-dust-mass ratio of 100. Four different gas-phase oxygen
  abundances were explore: the standard value Std, half of the
  standard value ([O]/2), one fourth ([O]/4), and one tenth
  ([O]/10).}\label{Low_oxygen_abundance}
\end{center}
\end{table*}
\section{Discussion}~\label{diskussion}

We have explored a small grid of models varying the gas-to-dust-mass
ratio and the gas vertical opening angle. None of the models manage to
fit the \OIfs\ line at 63 micron. The gas-to-dust-mass ratio is
constrained by the CO $J$=3--2 line to be between 50 and 100. The
\CIIfs\ line is sensitive to the opening angle and sets a limit of
10\%. Likewise, an opening angle of 10\% allows a better fit to the CO
ro-vibrational lines.

\subsection{A flat well-mixed disk around
  \source}~\label{diskussion_flat_disk}

All the line fluxes suggest that the disk is flat with an opening
angle smaller than 10\%. It is difficult for a vertical hydrostatic
disk to maintain such a flat disk for the gas. On the other hand, dust
grains can decouple from the gas and settle toward the midplane,
resulting in a flat dust disk. In a five-million-years old disk, most
of the dust grains are expected to have settled toward the
midplane. This is not clear for PAHs. A way to explain a flat
gas-and-dust disk would have been that the main heating agent of the
gas (PAHs) would also have settled toward the midplane. Without PAH
heating, a cool gas cannot sustain itself vertically and would have
collapsed toward the midplane.

\subsection{Gas and dust around
  \source}~\label{diskussion_gas_dust}

The total solid mass derived from the fit to the SED is $4.9\times
10^{-6}$ M$_\odot$, similar to the mass found by
\citet{Li2003ApJ...594..987L}, even though we adopted a much simpler
dust-grain composition and structure. Silicate dust grains are
affected by radiative pressure (RP) and Poynting-Robertson  (PR)
effects. \citet{Li2003ApJ...594..987L} estimated that $\sim$10$^{-4}$
M$_\odot$ of solids have been lost either by Poynting-Robertson
effect or by radiative pressure during the lifetime of the disk.
At least some of the currently observed dust grains have to be
replenished by collisional destruction of large
planetesimals. Assuming an initial gas-to-dust-mass ratio of 100, the
initial gas disk mass would have reached $\sim$10$^{-2}$ M$_\odot$.

We found a total PAH mass of 1.6~$\times$~10$^{-10}$ M$_\odot$,
whereas \citet{Li2003ApJ...594..987L} found a much lower PAH mass of
2.2~$\times$~10$^{-11}$ M$_\odot$. The difference may be ascribed to
the treatment of radiative transfer for the PAH excitation. The PAHs
can self-shield against the UV exciting photons. The
circumcircumcoronene is stable in the entire disk against
photodissociation. However, PAHs are subject to RP, PR, and
photodissociation. The ratio of radiative pressure to gravity is
$\beta_{\mathrm{RP}}\simeq$87.4 and is weakly size-dependent
\citep{Li2003ApJ...594..987L}. The total RP and PR PAH mass-loss rate
has been estimated to reach 9.8~$\times$~10$^{-14}$ M$_\odot$
yr$^{-1}$. In five million years, the \source\ disk would have lost
4.7~$\times$~10$^{-7}$ M$_\odot$ of PAHs. Assuming an initial
interstellar abundance relative to hydrogen of 3~$\times$~10$^{-7}$
for the PAHs \citep{Tielens2008ARA&A..46..289T}, the minimum initial
gas disk mass would have been 2~$\times$~10$^{-3}$ M$_\odot$. These
estimates do not take the drag between the gas and the PAHs into
account. These estimates do not include PAH destruction by
photodissociation, which mostly affects the small PAHs
\citep{Visser2007A&A...466..229V}. Therefore the current large PAHs in
the \source\ disk are those that have been stable against
photodissociation for 5~Myr.

The fine-structure and CO $J$=3--2 lines suggest that the \source\
disk currently has a gas mass of (2.5-5)~$\times$~10$^{-4}$ M$_\odot$,
which translates into a gas-to-dust-mass ratio of 50--100, consistent
with the interstellar value if more weight is given to the
fine-structure line constraints. On the other hand, a fit to the CO
$J$=3--2 flux alone by \citet{Dent2005MNRAS.359..663D} suggests a gas
mass of 5~$\times$~10$^{-5}$ M$_\odot$.
\citet{Jonkheid2006A&A...453..163J} modeled the outer disk around
\source\ with a gas mass of 2.4~$\times$~10$^{-4}$ M$_\odot$,
compatible with our derived gas mass range.

Rotational CO lines in disks tend to require a much lower disk mass
than from other gas-mass tracers
\citep{Bergin2013Natur.493..644B,Thi2001ApJ...561.1074T}.  Carbon
monoxide molecules can either be photodissociated or frozen onto grain
surfaces. The dust grain temperature is higher than 20~K throughout
the disk such that CO freeze-out is unlikely. In addition, the disk is
weakly optically thick in the UV range such that any CO molecule
adsorbed onto grain surfaces will be photodesorbed almost
immediately. The CO photodissociation and self-shielding processes are
complicated \citep{Visser2009A&A...503..323V}. Self-shielding effects
in 2D are treated approximately in the code. This simplification may
result in an overestimate of the CO abundance. Once
  photodissociated, carbon can remain in its neutral atomic form or can
  be photoionized. Detections of the two [CI] fine-structure emission
  lines in addition to the existing CO and \CII\ lines are necessary
  for determining the total gas-phase carbon budget in the disk assuming
  that \CII, \ion{C}{I}, and CO are the main gas-phase carbon-bearing
  species.
  
  The current gas mass is 100 to 1000 times lower than the initial gas
  disk mass estimated from the limited lifetime of the silicate dust
  grains and PAHs. In five million years, a mass-loss rate via
  accretion or photoevaporative wind of a few 10$^{-10}$-10$^{-9}$
  M$_\odot$ yr$^{-1}$ would have been sufficient for the disk to reach
  its current mass from an initially massive one. However, it is not
  clear whether a non-accreting, low X-ray emitter B9.5V star such as
  \source\ can provide enough ionizing photons to sustain a currently
  strong photoevaporative wind. If the dust and PAH loss-rates are
  correct, the \source\ disk should have been much more massive at its
  formation, with an initial estimated gas mass of at least 0.01
  M$_\odot$.

\section{Conclusion}\label{conclusion}

The {\it Herschel-PACS} spectral observations were used to constrain
the dust- and gas properties surrounding the 5-Myr old HerbigAe star
\source. The fit to the SED and mid-IR spectrum constrains the PAH
mass in the disk. The PAH abundance is depleted compared with the
interstellar value (3$\times$10$^{-7}$) by a factor
2$\times$10$^{-3}$ to 6.7$\times$10$^{-2}$. The gas emission lines are
best explained by a flat non-flaring disk. Most of the line fluxes are
reproduced within a factor two except for the \OIfs\ line at
63~$\mu$m.

We estimated the gas mass to be between 2 $\times$ 10$^{-4}$ and 4.9
$\times$ 10$^{-4}$ M$_\odot$ compared with the dust mass
($a_{\mathrm{max}}<$~1~mm) of 2.1 $\times$ 10$^{-6}$ M$_\odot$ or a
total solid mass ($a_{\mathrm{max}}$=1 cm) of 4.9 $\times$
10$^{-6}$ M$_\odot$. The large uncertainty in the disk gas mass
estimates comes most likely from our incomplete understanding of the
physical and chemical processes that determine the chemistry in
disks. Lowering the oxygen elemental abundance does not solve the
  problem of the very high \OIfs\ 63 micron line fluxes in our models.

\source\ is an example where the disk gas mass around a transitional
HerbigAe star has been constrained directly from gas phase
lines. However, the gas-to-dust-mass ratio depends on the gas tracer
used to derive its value. 

The disk solid and gas mass, as well as the disk scale-height are
lower than found around other Herbig~Ae stars in the {\it GASPS}
sample, but the gas-to-dust-mass ratio remains close to the initial
interstellar value of 100. If the disk around \source\ has initially
been massive ($\sim$~10$^{-2}$ M$_{\odot}$), the dissipation
mechanisms would have removed the gas and the dust simultaneously.

\begin{acknowledgements}
  We thank ANR (contracts ANR-07-BLAN-0221 and ANR-2010-JCJC-0504-01)
  and PNPS of CNRS/INSU, France for support. WFT, IK, and PW
  acknowledge funding from the EU FP7-2011 under Grant Agreement
  nr. 284405 (PERG06-GA-2009-256513).  FM acknowledges support from
  the Millenium Science Initiative (Chilean Ministry of Economy),
  through grant ''Nucleus P10-022-F''. Computations presented in this
  paper were performed at the Service Commun de Calcul Intensif de
  l'Observatoire de Grenoble (SCCI) on the super-computer Fostino
  funded by Agence Nationale pour la Recherche under contracts
  ANR-07-BLAN-0221, ANR-2010-JCJC-0504-01 and
  ANR-2010-JCJC-0501-01. C. Eiroa, G. Meeus, and B. Montesinos are
  partly supported by Spanish grant AYA 2011-26202. We thank the
    referee for the useful comments.
\end{acknowledgements}
  
\bibliographystyle{aa}
\bibliography{hd141569_bib}

\begin{appendix}

\section{Appendix}

\subsection{PACS Spatial pixel emissions}\label{spaxels}

\begin{figure*}[!ht]  
\resizebox{\hsize}{!}{\includegraphics[scale=1,angle=90]{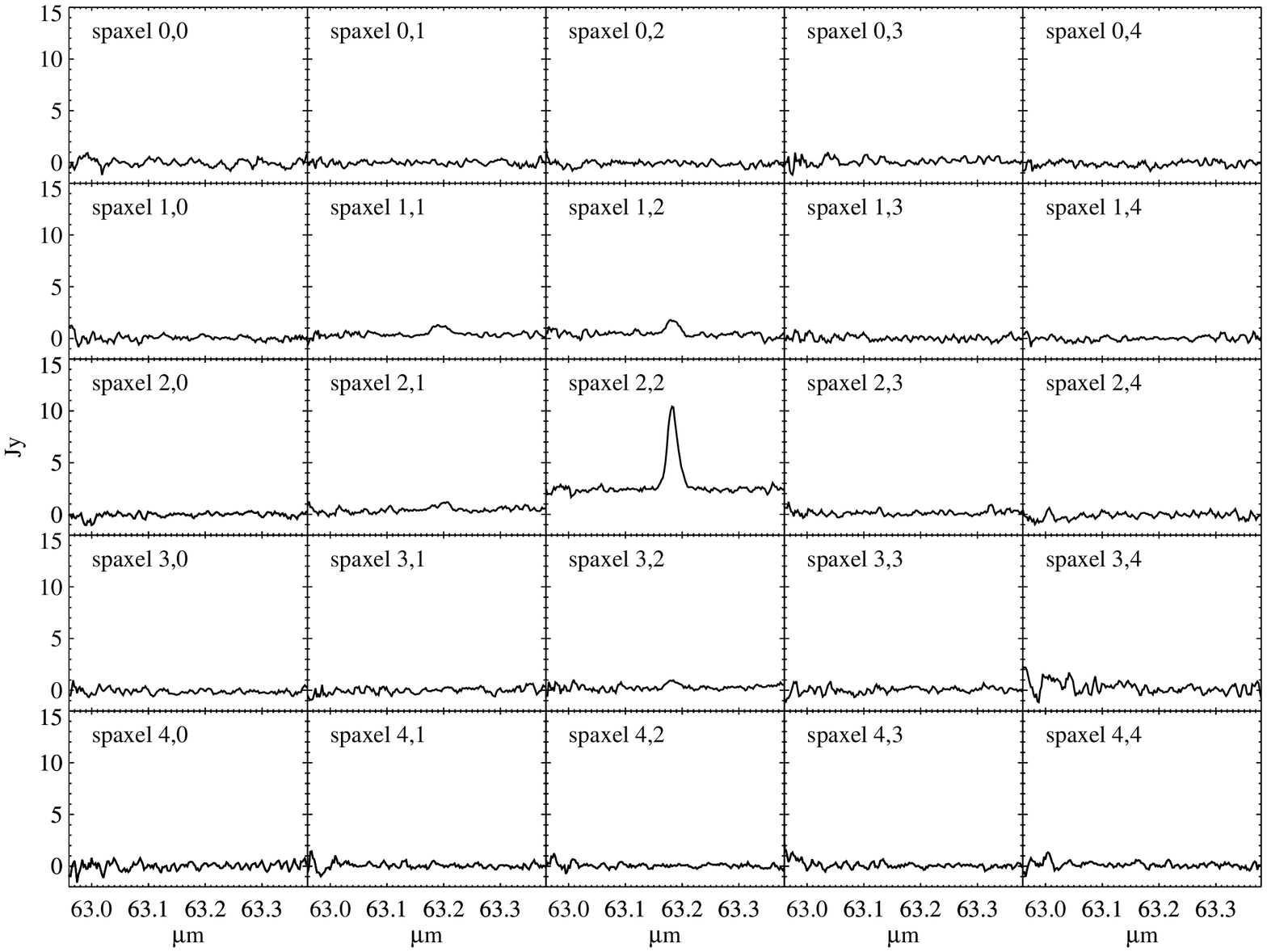}}  
\caption{\OIfs\ 63 $\mu$m spectrum for each pixel
  in the PACS spectrometer array. The PACS integral filed unit has a
  47\arcsec $\times$ 47\arcsec field of view. Each 25 pixels has a
  9.4\arcsec $\times$ 9.4\arcsec spatial resolution. The flux is
  emitted in the central spaxels.}\label{OI63_spaxels}
\end{figure*}  
\begin{figure*}[!ht]  
\resizebox{\hsize}{!}{\includegraphics[scale=1,angle=90]{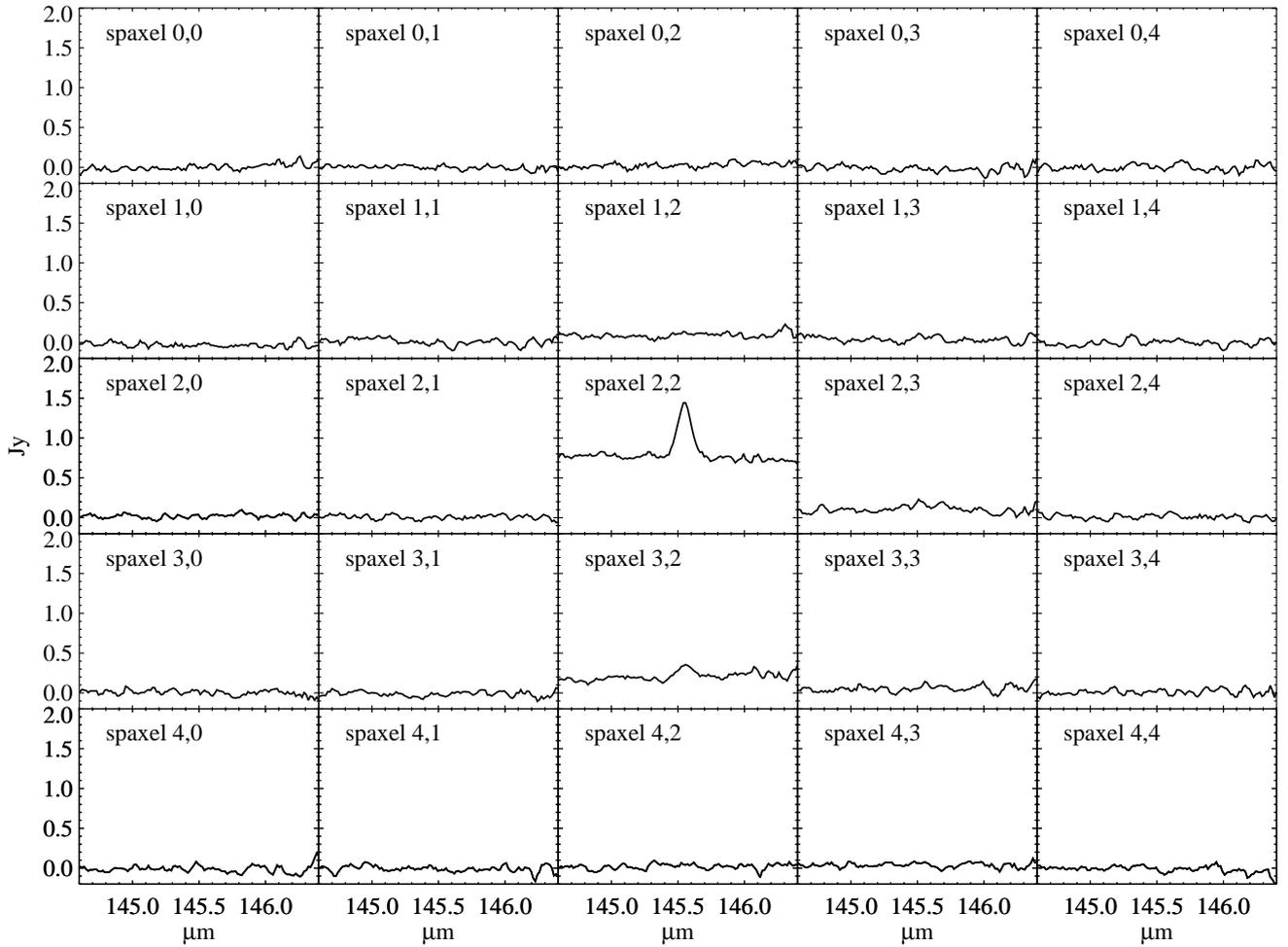}}  
\caption{\OIfs\ 145 $\mu$m spectrum for each pixel
  in the PACS spectrometer array.}\label{OI145_spaxels}
\end{figure*}  
\begin{figure*}[!ht]  
\resizebox{\hsize}{!}{\includegraphics[scale=1,angle=90]{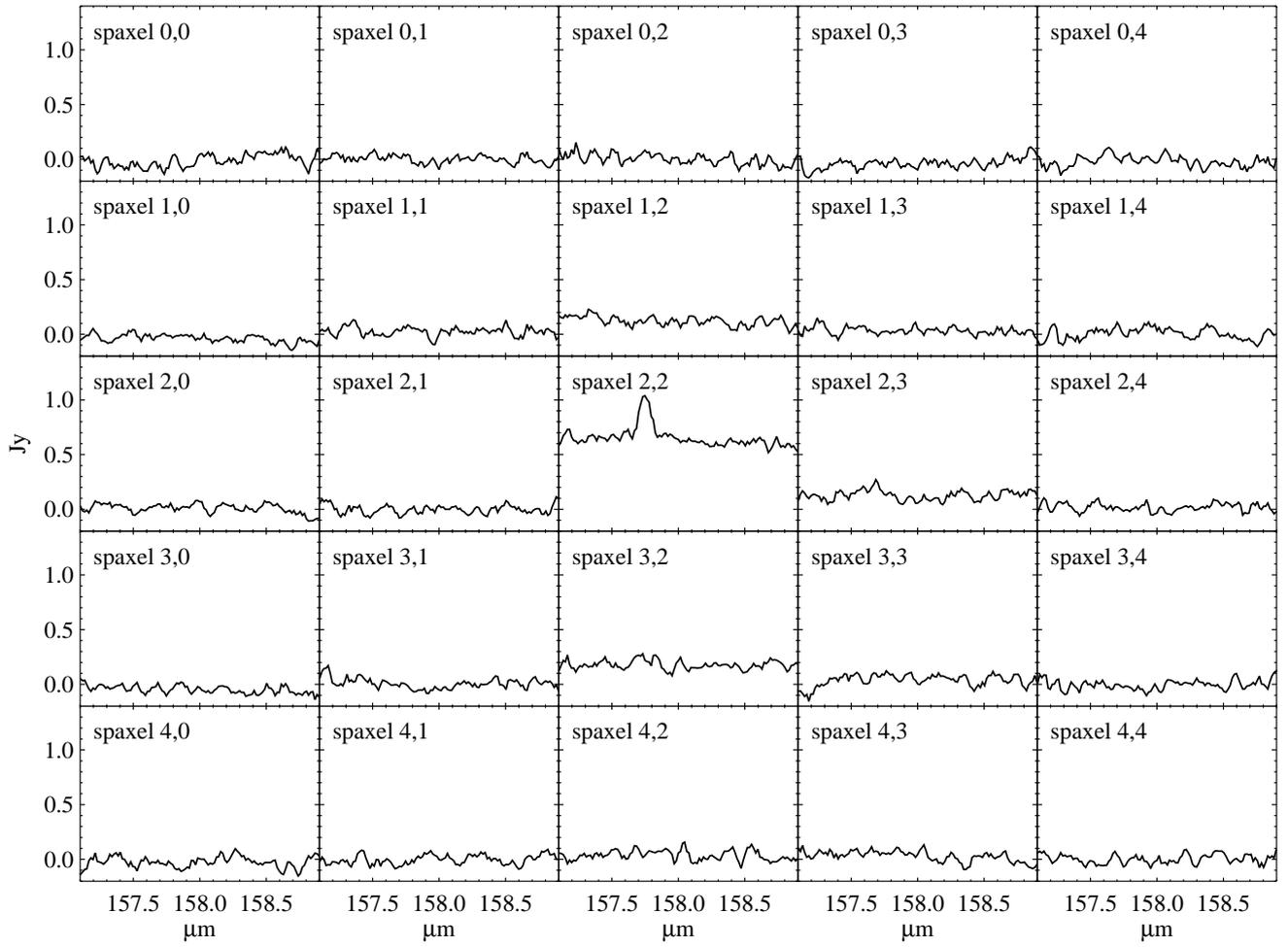}}  
\caption{\CIIfs\ spectrum for each pixel in the
  PACS spectrometer array. The \CIIfs\ 157$\mu$m line is detected in
  the central spaxel.}\label{CII_spaxels}
\end{figure*}  

\subsection{Collisional data}\label{collisional_rates}

The original articles for the line frequencies, Einstein coefficients,
and collisional rates are CO
\citep{Flower2001JPhB...34.2731F,Jankowski2005JChPh.123j4301J,Yang2006JChPh.124j4304Y,Wernli2006A&A...446..367W},
H$_2$O
\citep{Barber2006MNRAS.368.1087B,Dubernet2002A&A...390..793D,Faure2004MNRAS.347..323F,Faure2007A&A...472.1029F,Daniel2011A&A...536A..76D},
\OI\
\citep{Abrahamsson2007ApJ...654.1171A,Bell1998MNRAS.293L..83B,Chambaud1980JPhB...13.4205C,Jaquet1992JPhB...25..285J,Launay1977JPhB...10..879L},
\CII\
\citep{Flower1977JPhB...10.3673F,Launay1977JPhB...10..879L,Wilson2002MNRAS.337.1027W},
CH$^+$
\citep{Muller2010A&A...514L...6M,Lim1999MNRAS.306..473L,Hammami2009A&A...507.1083H,turpin2010A&A...511A..28T},
OH \citep{Offer1994JChPh.100..362O}.

\subsection{Photometric data}

We provide the photometric data used in our SED modeling in
Table~\ref{table_photometry}.

\begin{table}
\centering
\caption{Non-simultaneous photometric data. The data without reference are taken from \citet{Merin2004A&A...419..301M}.}
\label{table_photometry} 
\begin{tabular}{llll}   
  \hline 
  \noalign{\smallskip}   
  \multicolumn{1}{c}{Band} &  \multicolumn{1}{c}{$\lambda$} & \multicolumn{1}{c}{Flux} & \multicolumn{1}{c}{(Beam size) and reference}\\
  &  \multicolumn{1}{c}{($\mu$m)}  &  \multicolumn{1}{c}{(Jy)}& \\
  \noalign{\smallskip} 
  \hline
  \noalign{\smallskip} 
  {\it IUE} & 0.138 & 0.30  & archival data\\
  {\it IUE} & 0.178 & 0.73  & archival data \\
  {\it IUE} & 0.218 & 0.628 & archival data\\
  {\it IUE} & 0.257 & 1.023 & archival data \\
  {\it IUE} & 0.29  & 1.274 & archival data \\
  {\it U}   & 0.36  & 4.19  & {\citet{Sylvester1996MNRAS.279..915S}} \\
  {\it B}   & 0.436 & 8.37  & {\citet{Sylvester1996MNRAS.279..915S}}\\
  {\it V}   & 0.55  & 7.36  & {\citet{Sylvester1996MNRAS.279..915S}} \\
  {\it R}   & 0.708 & 5.92  & {\citet{Sylvester1996MNRAS.279..915S}} \\
  {\it I}   & 0.977 & 4.81  & {\citet{Sylvester1996MNRAS.279..915S}} \\
  {\it J}   & 1.22  & 3.1   &  2Mass\\
  {\it H}   & 1.65  & 1.8   &  2Mass\\
  {\it K}   & 2.18  & 1.2   &  2Mass\\  
  {\it ISO} & 2.45  & 1.07  &  ESA archive \\
  {\it ISO} & 3.23  & 0.76  &  ESA archive\\
  {\it WISE} & 3.4  & 0.79~$\pm$~0.025 & 6.1 $\arcsec$ NASA archive\\ 
  {\it UKIRT} & 3.76  & 0.54  &   {\citet{Sylvester1996MNRAS.279..915S}}\\
  {\it ISO}  & 4.26  & 0.44  & ESA archive  \\
  {\it WISE} & 4.6   & 0.49~$\pm$~0.01 & 6.4 $\arcsec$ \\ 
  {\it ISO}  & 5.89  & 0.40  & ESA archive  \\
  {\it ISO}  & 6.76  & 0.43  & ESA archive  \\  
  {\it ISO}  & 7.76  & 0.82  & ESA archive \\
  {\it ISO}  & 8.70  & 0.62  & ESA archive  \\
  {\it AKARI} & 9.0  & 0.5178~$\pm$~0.0104 & NASA archive\\
  {\it ISO}  & 9.77  & 0.52  & ESA archive  \\
  {\it ISO}  & 10.7  & 0.58  & ESA archive  \\
  {\it OSCIR} & 10.8  & 0.318~$\pm$~0.016 & {\citet{Fisher2000ApJ...532L.141F}}\\
  {\it Michelle} & 11.2 & 0.338~$\pm$~0.034 & {\citet{Moerchen2010ApJ...723.1418M}}\\
  {\it ISO}           & 11.48 & 0.635 & 14$\arcsec$ $\times$ 20 $\arcsec$ ESA archive\\
  {\it IRAS} & 12.0  & 0.55~$\pm$~0.04 & 1$'$ $\times$ 5$'$ NASA archive\\
  {\it WISE} & 12.0  & 0.38~$\pm$~0.006 & 6.5 $\arcsec$ NASA archive\\ 
  {\it MIRLIN}   & 12.5  & 0.333~$\pm$~0.022 & {\citet{Marsh2002ApJ...573..425M}} \\
  {\it MIRLIN}   & 17.9  & 0.936~$\pm$~0.094 & {\citet{Marsh2002ApJ...573..425M}} \\
  {\it AKARI}    & 18 & 0.8655~$\pm$~0.0168 & NASA archive\\ 
  {\it Michelle} & 18.1 & 0.883~$\pm$~0.147 & {\citet{Moerchen2010ApJ...723.1418M}} \\ 
  {\it OSCIR} & 18.2  & 0.646~$\pm$~0.035 & {\citet{Fisher2000ApJ...532L.141F}}\\
  {\it MIRLIN} & 20.8  & 1.19~$\pm$~0.16 & {\citet{Marsh2002ApJ...573..425M}} \\ 
  {\it WISE} & 22 & 1.44~$\pm$~0.027 & 12$\arcsec$ NASA archive\\ 
  {\it MIPS} & 24.0  & 1.47~$\pm$~0.01 & 6~$\arcsec$ Spitzer archive\\
  {\it IRAS}  & 25   & 1.87~$\pm$~0.13 & 1$'$ $\times$ 5$'$ NASA archive\\  
  {\it IRAS}  & 60   & 5.54~$\pm$~0.49 & 2$'$ $\times$ 5$'$ NASA archive\\
  {\it PACS-Spec} & 63.2& 2.98~$\pm$~0.01 & this paper\\
  {\it MIPS} & 70 & 4.70~$\pm$~0.02 & 18~$\arcsec$ Spitzer archive\\ 
  {\it PACS-Spec} & 72.8 & 3.91~$\pm$~0.03 & this paper\\
  {\it PACS-Spec} & 76.4 & 3.30~$\pm$~0.03 & this paper\\
  {\it PACS-Spec} & 90 & 2.80~$\pm$~0.03 & this paper\\
  {\it IRAS}  &  100  & 3.48~$\pm$~0.35 & 4$'$ $\times$ 5$'$ NASA archive\\
  {\it PACS-Spec} & 145 & 1.1~$\pm$~0.1 & this paper\\
  {\it PACS-Spec} & 158 & 1.18~$\pm$~0.02 & this paper\\
  {\it PACS-Spec} & 180 & 0.83~$\pm$~0.04 & this paper\\
  {\it SCUBA} &  450  & 0.0649~$\pm$~0.0133 & {\citet{Sandell2011ApJ...727...26S}}\\
  {\it SCUBA} &  850  & 0.0140~$\pm$~0.0020 & {\citet{Sandell2011ApJ...727...26S}}\\
  {\it LABOCA}  &  870  & 0.0126~$\pm$~0.0046 & {\citet{Nilsson2010A&A...518A..40N}}\\
  {\it MAMBO} & 1200 & 0.0047~$\pm$~0.0005 & {\citet{Meeus2012A&A...544A..78M}}\\
  {\it SCUBA}   & 1350  & 0.0054~$\pm$~0.0001 & {\citet{Sylvester2001MNRAS.327..133S}}\\   
  \noalign{\smallskip} 
  \hline
\end{tabular}
\end{table}

\end{appendix}  

\end{document}